\documentclass[11pt, a4paper]{article}
\PassOptionsToPackage{unicode}{hyperref}
\usepackage{amssymb}
\usepackage{jheppub}
\usepackage{upgreek}
\usepackage{slashed}

\title{Poisson vertex algebras in supersymmetric field theories}

\author[1,2]{Jihwan Oh}
\author[1]{and Junya Yagi}

\affiliation[1]{Perimeter Institute for Theoretical Physics,
  Waterloo, ON
  N2L 2Y5 Canada}
\affiliation[2]{Department of Physics,
  University of California, Berkeley, CA 94720 USA}

\abstract{A large class of supersymmetric quantum field theories,
  including all theories with $\CN = 2$ supersymmetry in three
  dimensions and theories with $\CN = 2$ supersymmetry in four
  dimensions, possess topological--holomorphic sectors.  We formulate
  Poisson vertex algebras in such topological--holomorphic sectors and
  discuss some examples.  For a four-dimensional $\CN = 2$
  superconformal field theory, the associated Poisson vertex algebra
  is the classical limit of a vertex algebra generated by a subset of
  local operators of the theory.}


\arxivnumber{}

\let\U\relax
\let\C\relax

\newcommand{\gf}{\mathfrak{g}}

\newcommand{\del}{\partial}
\newcommand{\delb}{{\bar\partial}}

\newcommand{\id}{\mathop{\mathrm{id}}\nolimits}

\def\ie{\begin{equation}\begin{aligned}}
\def\fe{\end{aligned}\end{equation}}

\newcommand{\Tr}{\mathop{\mathrm{Tr}}\nolimits}

\newcommand{\SU}{\mathrm{SU}}

\newcommand{\SL}{\mathrm{SL}}

\newcommand{\U}{\mathrm{U}}

\newcommand{\iso}{\cong}

\newcommand{\Z}{\mathbb{Z}}

\newcommand{\R}{\mathbb{R}}
\newcommand{\C}{\mathbb{C}}

\let\nc\newcommand
\let\renc\renewcommand
\nc{\wbar}{\overline}
\let\td\tilde
\let\wtd\widetilde
\let\wht\widehat
\let\mcl\mathcal

\nc{\ab}{{\bar{a}}} \nc{\at}{\tilde{a}} \nc{\ah}{\hat{a}}
\nc{\bb}{{\bar{b}}} \nc{\bt}{\tilde{b}} \nc{\bh}{\hat{b}}
\nc{\cb}{{\bar{c}}} \nc{\ct}{\tilde{c}} %\nc{\ch}{\hat{c}}
\nc{\db}{{\bar{d}}} \nc{\dt}{\tilde{d}} \renc{\dh}{\hat{d}}
\nc{\eb}{{\bar{e}}} \nc{\et}{\tilde{e}} \nc{\eh}{\hat{e}}
\nc{\fb}{{\bar{f}}} \nc{\ft}{\tilde{f}} \nc{\fh}{\hat{f}}
\nc{\gb}{{\bar{g}}} \nc{\gt}{\tilde{g}} \nc{\gh}{\hat{g}}
\nc{\hb}{{\bar{h}}} \nc{\hh}{\hat{h}} %\nc{\ht}{\tilde{h}}
\nc{\ib}{{\bar{\imath}}} \nc{\ih}{\hat{\imath}} %\nc{\it}{\tilde{\imath}}
\nc{\jb}{{\bar{\jmath}}} \nc{\jt}{\tilde{\jmath}} \nc{\jh}{\hat{\jmath}}
\nc{\kb}{{\bar{k}}} \nc{\kt}{\tilde{k}} \nc{\kh}{\hat{k}}
\nc{\lb}{{\bar{l}}} \nc{\lt}{\tilde{l}} \nc{\lh}{\hat{l}}
\nc{\mb}{{\bar{m}}} \nc{\mt}{\tilde{m}} \nc{\mh}{\hat{m}}
\nc{\nb}{{\bar{n}}} \nc{\nt}{\tilde{n}} \nc{\nh}{\hat{n}}
\nc{\ob}{{\bar{o}}} \nc{\ot}{\tilde{o}} \nc{\oh}{\hat{o}}
\nc{\pb}{{\bar{p}}} \nc{\pt}{\tilde{p}} \nc{\ph}{\hat{p}}
\nc{\qb}{{\bar{q}}} \nc{\qt}{\tilde{q}} \nc{\qh}{\hat{q}}
\nc{\rb}{{\bar{r}}} \nc{\rt}{\tilde{r}} \nc{\rh}{{\hat{r}}}
\renc{\sb}{{\bar{s}}} \nc{\st}{\tilde{s}} \nc{\sh}{\hat{s}}
\nc{\tb}{{\bar{t}}} \renc{\th}{\hat{t}} %\nc{\tt}{\tilde{t}}
\nc{\ub}{{\bar{u}}} \nc{\ut}{\tilde{u}} \nc{\uh}{\hat{u}}
\nc{\vb}{{\bar{v}}} \nc{\vt}{\tilde{v}} \nc{\vh}{\hat{v}}
\nc{\wb}{{\bar{w}}} \nc{\wt}{\tilde{w}} \nc{\wh}{\hat{w}}
\nc{\xb}{{\bar{x}}} \nc{\xt}{\tilde{x}} \nc{\xh}{\hat{x}}
\nc{\yb}{{\bar{y}}} \nc{\yt}{\tilde{y}} \nc{\yh}{\hat{y}}
\nc{\zb}{{\bar{z}}} \nc{\zt}{\tilde{z}} \nc{\zh}{\hat{z}}

\nc{\Ab}{{\wbar{A}}} \nc{\At}{{\wtd{A}}} \nc{\Ah}{{\wht{A}}}
\nc{\Bb}{{\wbar{B}}} \nc{\Bt}{{\wtd{B}}} \nc{\Bh}{{\wht{B}}}
\nc{\Cb}{{\wbar{C}}} \nc{\Ct}{{\wtd{C}}} \nc{\Ch}{{\wht{C}}}
\nc{\Db}{{\wbar{D}}} \nc{\Dt}{{\wtd{D}}} \nc{\Dh}{{\wht{D}}}
\nc{\Eb}{{\wbar{E}}} \nc{\Et}{{\wtd{E}}} \nc{\Eh}{{\wht{E}}}
\nc{\Fb}{{\wbar{F}}} \nc{\Ft}{{\wtd{F}}} \nc{\Fh}{{\wht{F}}}
\nc{\Gb}{{\wbar{G}}} \nc{\Gt}{{\wtd{G}}} \nc{\Gh}{{\wht{G}}}
\nc{\Hb}{{\wbar{H}}} \nc{\Ht}{{\wtd{H}}} \nc{\Hh}{{\wht{H}}}
\nc{\Ib}{{\bar{I}}} \nc{\It}{{\wtd{I}}} \nc{\Ih}{{\wht{I}}}
\nc{\Jb}{{\wbar{J}}} \nc{\Jt}{{\wtd{J}}} \nc{\Jh}{{\wht{J}}}
\nc{\Kb}{{\wbar{K}}} \nc{\Kt}{{\wtd{K}}} \nc{\Kh}{{\wht{K}}}
\nc{\Lb}{{\wbar{L}}} \nc{\Lt}{{\wtd{L}}} \nc{\Lh}{{\wht{L}}}
\nc{\Mb}{{\wbar{M}}} \nc{\Mt}{{\wtd{M}}} \nc{\Mh}{{\wht{M}}}
\nc{\Nb}{{\wbar{N}}} \nc{\Nt}{{\wtd{N}}} \nc{\Nh}{{\wht{N}}}
\nc{\Ob}{{\wbar{O}}} \nc{\Ot}{{\wtd{O}}} \nc{\Oh}{{\wht{O}}}
\nc{\Pb}{{\wbar{P}}} \nc{\Pt}{{\wtd{P}}} \nc{\Ph}{{\wht{P}}}
\nc{\Qb}{{\wbar{Q}}} \nc{\Qt}{{\wtd{Q}}} \nc{\Qh}{{\wht{Q}}}
\nc{\Rb}{{\wbar{R}}} \nc{\Rt}{{\wtd{R}}} \nc{\Rh}{{\wht{R}}}
\nc{\Sb}{{\wbar{S}}} \nc{\St}{{\wtd{S}}} \nc{\Sh}{{\wht{S}}}
\nc{\Tb}{{\wbar{T}}} \nc{\Tt}{{\wtd{T}}} \nc{\Th}{{\wht{T}}}
\nc{\Ub}{{\wbar{U}}} \nc{\Ut}{{\wtd{U}}} \nc{\Uh}{{\wht{U}}}
\nc{\Vb}{{\wbar{V}}} \nc{\Vt}{{\wtd{V}}} \nc{\Vh}{{\wht{V}}}
\nc{\Wb}{{\wbar{W}}} \nc{\Wt}{{\wtd{W}}} \nc{\Wh}{{\wht{W}}}
\nc{\Xb}{{\wbar{X}}} \nc{\Xt}{{\wtd{X}}} \nc{\Xh}{{\wht{X}}}
\nc{\Yb}{{\wbar{Y}}} \nc{\Yt}{{\wtd{Y}}} \nc{\Yh}{{\wht{Y}}}
\nc{\Zb}{{\wbar{Z}}} \nc{\Zt}{{\wtd{Z}}} \nc{\Zh}{{\wht{Z}}}

\nc{\CA}{{\mcl{A}}} \nc{\CAb}{{\wbar{\CA}}} \nc{\CAt}{{\wtd{\CA}}} \nc{\CAh}{{\wht{\CA}}}
\nc{\CB}{{\mcl{B}}} \nc{\CBb}{{\wbar{\CB}}} \nc{\CBt}{{\wtd{\CB}}} \nc{\CBh}{{\wht{\CB}}}
\nc{\CC}{{\mcl{C}}} \nc{\CCb}{{\wbar{\CC}}} \nc{\CCt}{{\wtd{\CC}}} \nc{\CCh}{{\wht{\CC}}}
\nc{\cD}{{\mcl{D}}} \nc{\cDb}{{\wbar{\cD}}} \nc{\cDt}{{\wtd{\cC}}} \nc{\cDh}{{\wht{\cD}}}
\nc{\CE}{{\mcl{E}}} \nc{\CEb}{{\wbar{\CE}}} \nc{\CEt}{{\wtd{\CE}}} \nc{\CEh}{{\wht{\CE}}}
\nc{\CF}{{\mcl{F}}} \nc{\CFb}{{\wbar{\CF}}} \nc{\CFt}{{\wtd{\CF}}} \nc{\CFh}{{\wht{\CF}}}
\nc{\CG}{{\mcl{G}}} \nc{\CGb}{{\wbar{\CG}}} \nc{\CGt}{{\wtd{\CG}}} \nc{\CGh}{{\wht{\CG}}}
\nc{\CH}{{\mcl{H}}} \nc{\CHb}{{\wbar{\CH}}} \nc{\CHt}{{\wtd{\CH}}} \nc{\CHh}{{\wht{\CH}}}
\nc{\CI}{{\mcl{I}}} \nc{\CIb}{{\wbar{\CI}}} \nc{\CIt}{{\wtd{\CI}}} \nc{\CIh}{{\wht{\CI}}}
\nc{\CJ}{{\mcl{J}}} \nc{\CJb}{{\wbar{\CJ}}} \nc{\CJt}{{\wtd{\CJ}}} \nc{\CJh}{{\wht{\CJ}}}
\nc{\CK}{{\mcl{K}}} \nc{\CKb}{{\wbar{\CK}}} \nc{\CKt}{{\wtd{\CK}}} \nc{\CKh}{{\wht{\CK}}}
\nc{\CL}{{\mcl{L}}} \nc{\CLb}{{\wbar{\CL}}} \nc{\CLt}{{\wtd{\CL}}} \nc{\CLh}{{\wht{\CL}}}
\nc{\CM}{{\mcl{M}}} \nc{\CMb}{{\wbar{\CM}}} \nc{\CMt}{{\wtd{\CM}}} \nc{\CMh}{{\wht{\CM}}}
\nc{\CN}{{\mcl{N}}} \nc{\CNb}{{\wbar{\CN}}} \nc{\CNt}{{\wtd{\CN}}} \nc{\CNh}{{\wht{\CN}}}
\nc{\CO}{{\mcl{O}}} \nc{\COb}{{\wbar{\CO}}} \nc{\COt}{{\wtd{\CO}}} \nc{\COh}{{\wht{\CO}}}
\nc{\CP}{{\mcl{P}}} \nc{\CPb}{{\wbar{\CP}}} \nc{\CPt}{{\wtd{\CP}}} \nc{\CPh}{{\wht{\CP}}}
\nc{\CQ}{{\mcl{Q}}} \nc{\CQb}{{\wbar{\CQ}}} \nc{\CQt}{{\wtd{\CQ}}} \nc{\CQh}{{\wht{\CQ}}}
\nc{\CR}{{\mcl{R}}} \nc{\CRb}{{\wbar{\CR}}} \nc{\CRt}{{\wtd{\CR}}} \nc{\CRh}{{\wht{\CR}}}
\nc{\CS}{{\mcl{S}}} \nc{\CSb}{{\wbar{\CS}}} \nc{\CSt}{{\wtd{\CS}}} \nc{\CSh}{{\wht{\CS}}}
\nc{\CT}{{\mcl{T}}} \nc{\CTb}{{\wbar{\CT}}} \nc{\CTt}{{\wtd{\CT}}} \nc{\CTh}{{\wht{\CT}}}
\nc{\CU}{{\mcl{U}}} \nc{\CUb}{{\wbar{\CU}}} \nc{\CUt}{{\wtd{\CU}}} \nc{\CUh}{{\wht{\CU}}}
\nc{\CV}{{\mcl{V}}} \nc{\CVb}{{\wbar{\CV}}} \nc{\CVt}{{\wtd{\CV}}} \nc{\CVh}{{\wht{\CV}}}
\nc{\CW}{{\mcl{W}}} \nc{\CWb}{{\wbar{\CW}}} \nc{\CWt}{{\wtd{\CW}}} \nc{\CWh}{{\wht{\CW}}}
\nc{\CX}{{\mcl{X}}} \nc{\CXb}{{\wbar{\CX}}} \nc{\CXt}{{\wtd{\CX}}} \nc{\CXh}{{\wht{\CX}}}
\nc{\CY}{{\mcl{Y}}} \nc{\CYb}{{\wbar{\CY}}} \nc{\CYt}{{\wtd{\CY}}} \nc{\CYh}{{\wht{\CY}}}
\nc{\CZ}{{\mcl{Z}}} \nc{\CZb}{{\wbar{\CZ}}} \nc{\CZt}{{\wtd{\CZ}}} \nc{\CZh}{{\wht{\CZ}}}

\let\eps\epsilon
\let\ups\upsilon
\let\veps\varepsilon
\let\vtht\vartheta

\let\vsgm\varsigma
\let\vphi\varphi
\let\vrho\varrho

\nc{\alphab}{{\bar{\alpha}}} \nc{\alphat}{{\td{\alpha}}} \nc{\alphah}{{\hat{\alpha}}}
\nc{\betab}{{\bar{\beta}}}   \nc{\betat}{{\td{\beta}}}   \nc{\betah}{{\hat{\beta}}} 
\nc{\gammab}{{\bar{\gamma}}} \nc{\gammat}{{\td{\gamma}}} \nc{\gammah}{{\hat{\gamma}}} 
\nc{\deltab}{{\bar{\delta}}} \nc{\deltat}{{\td{\delta}}} \nc{\deltah}{{\hat{\delta}}} 
\nc{\epsilonb}{{\bar{\eps}}} \nc{\epsilont}{{\td{\eps}}} \nc{\epsilonh}{{\hat{\eps}}} 
\nc{\vepsb}{{\bar{\veps}}}   \nc{\vepst}{{\td{\veps}}}   \nc{\vepsh}{{\hat{\veps}}} 
\nc{\zetab}{{\bar{\zeta}}}   \nc{\zetat}{{\td{\zeta}}}   \nc{\zetah}{{\hat{\zeta}}} 
\nc{\etab}{{\bar{\eta}}}     \nc{\etat}{{\td{\eta}}}     \nc{\etah}{{\hat{\eta}}} 
\nc{\thetab}{{\bar{\theta}}} \nc{\thetat}{{\td{\theta}}} \nc{\thetah}{{\hat{\theta}}} 
\nc{\vthetab}{{\bar{\vtht}}} \nc{\vthetat}{{\td{\vtht}}} \nc{\vthetah}{{\hat{\vtht}}} 
\nc{\lambdab}{{\bar{\lambda}}} \nc{\lambdat}{{\td{\lambda}}} \nc{\lambdah}{{\hat{\lambda}}} 
\nc{\iotab}{{\bar{\iota}}}   \nc{\iotat}{{\td{\iota}}}   \nc{\iotah}{{\hat{\iota}}} 
\nc{\kappab}{{\bar{\kappa}}} \nc{\kappat}{{\td{\kappa}}} \nc{\kappah}{{\hat{\kappa}}} 
\nc{\lmdb}{{\bar{\lmd}}}     \nc{\lmdt}{{\td{\lmd}}}     \nc{\lmdh}{{\hat{\lmd}}} 
\nc{\mub}{{\bar{\mu}}}       \nc{\mut}{{\td{\mu}}}       \nc{\muh}{{\hat{\mu}}} 
\nc{\nub}{{\bar{\nu}}}       \nc{\nut}{{\td{\nu}}}       \nc{\nuh}{{\hat{\nu}}} 
\nc{\xib}{{\bar{\xi}}}       \nc{\xit}{{\td{\xi}}}       \nc{\xih}{{\hat{\xi}}} 
\nc{\pib}{{\bar{\pi}}}       \nc{\pit}{{\td{\pi}}}       \nc{\pih}{{\hat{\pi}}} 
\nc{\vpib}{{\bar{\vpi}}}     \nc{\vpit}{{\td{\vpi}}}     \nc{\vpih}{{\hat{\vpi}}} 
\nc{\rhob}{{\bar{\rho}}}     \nc{\rhot}{{\td{\rho}}}     \nc{\rhoh}{{\hat{\rho}}} 
\nc{\vrhob}{{\bar{\vrho}}}   \nc{\vrhot}{{\td{\vrho}}}   \nc{\vrhoh}{{\hat{\vrho}}} 
\nc{\sigmab}{{\bar{\sigma}}} \nc{\sigmat}{{\td{\sigma}}} \nc{\sigmah}{{\hat{\sigma}}} 
\nc{\vsigmab}{{\bar{\vsgm}}} \nc{\vsigmat}{{\td{\vsgm}}} \nc{\vsigmah}{{\hat{\vsgm}}} 
\nc{\taub}{{\bar{\tau}}}     \nc{\taut}{{\td{\tau}}}     \nc{\tauh}{{\hat{\tau}}} 
\nc{\upsb}{{\bar{\ups}}} \nc{\upst}{{\td{\ups}}} \nc{\upsh}{{\hat{\ups}}} 
\nc{\phib}{{\bar{\phi}}}     \nc{\phit}{{\td{\phi}}}     \nc{\phih}{{\hat{\phi}}} 
\nc{\varphib}{{\bar{\vphi}}}   \nc{\varphit}{{\td{\vphi}}}   \nc{\varphih}{{\hat{\vphi}}} 
\nc{\chib}{{\bar{\chi}}}     \nc{\chit}{{\td{\chi}}}     \nc{\chih}{{\hat{\chi}}} 
\nc{\psib}{{\bar{\psi}}}     \nc{\psit}{{\td{\psi}}}     \nc{\psih}{{\hat{\psi}}} 
\nc{\omegab}{{\bar{\omega}}} \nc{\omegat}{{\td{\omega}}} \nc{\omegah}{{\hat{\omega}}} 

\nc{\Gammab}{{\wbar{\Gamma}}}     \nc{\Gammat}{{\wtd{\Gamma}}}     \nc{\Gammah}{{\wht{\Gamma}}}
\nc{\Deltab}{{\wbar{\Delta}}}     \nc{\Deltat}{{\wtd{\Delta}}}     \nc{\Deltah}{{\wht{\Delta}}}
\nc{\Thetab}{{\wbar{\Theta}}}     \nc{\Thetat}{{\wtd{\Theta}}}     \nc{\Thetah}{{\wht{\Theta}}}
\nc{\Lambdab}{{\wbar{\Lambda}}}   \nc{\Lambdat}{{\wtd{\Lambda}}}   \nc{\Lambdah}{{\wht{\Lambda}}}
\nc{\Xib}{{\wbar{\Xi}}}           \nc{\Xit}{{\wtd{\Xi}}}           \nc{\Xih}{{\wht{\Xi}}}
\nc{\Pib}{{\wbar{\Pi}}}           \nc{\Pit}{{\wtd{\Pi}}}           \nc{\Pih}{{\wht{\Pi}}}
\nc{\Sigmab}{{\wbar{\Sigma}}}     \nc{\Sigmat}{{\wtd{\Sigma}}}     \nc{\Sigmah}{{\wht{\Sigma}}}
\nc{\Upsilonb}{{\wbar{\Upsilon}}} \nc{\Upsilont}{{\wtd{\Upsilon}}} \nc{\Upsilonh}{{\wht{\Upsilon}}}
\nc{\Phib}{{\wbar{\Phi}}} \nc{\Phit}{{\wtd{\Phi}}} \nc{\Phih}{{\wht{\Phi}}}
\nc{\Psib}{{\wbar{\Psi}}}         \nc{\Psit}{{\wtd{\Psi}}}         \nc{\Psih}{{\wht{\Psi}}}
\nc{\Omegab}{{\wbar{\Omega}}}     \nc{\Omegat}{{\wtd{\Omega}}}     \nc{\Omegah}{{\wht{\Omega}}}

\newcommand{\rmd}{\mathrm{d}}
\newcommand{\epsb}{\epsilonb}

\newcommand{\iu}{\mathrm{i}}

\let\starx\star
\let\star\relax
\newcommand{\star}{\mathop{\starx}\nolimits}

\newcommand{\BRST}{\text{BRST}}

\usepackage[mathscr]{euscript}

\newcommand{\Qone}{\mathbf{Q}}

\newcommand{\conf}{\mathrm{Conf}}

\newcommand{\lbracket}[3]{\{#2 {}_{\,#1\,} #3\}}
\newcommand{\lBracket}[3]{[#2 {}_{\,#1\,} #3]}

\newcommand{\zmode}{{\textstyle\int}}

\newcommand{\ec}[1]{[\![#1]\!]}

\newcommand{\biggec}[1]{\biggl[\!\!\biggl[#1\biggr]\!\!\biggr]}

\begin{document}
\maketitle

\section{Introduction}

In this work we study algebraic structures in topological--holomorphic
sectors of supersymmetric quantum field theories, with one or more
topological directions and a single holomorphic direction.  These
structures, known as \emph{Poisson vertex
  algebras}~\cite{MR2058353,MR1620526,MR2082709}, exist in $\CN = 2$
supersymmetric field theories in three dimensions and $\CN = 2$
supersymmetric field theories in four dimensions, among others.

Our motivation for studying the Poisson vertex algebras comes from two
main sources.

One is the fact that local operators in topological quantum field
theories (TQFTs) of cohomological
type~\cite{Witten:1988ze,Witten:1988xj} form Poisson algebras, some
aspects of which were elucidated in the recent
paper~\cite{Beem:2018fng}.  TQFTs being special cases of
topological--holomorphic theories, it is natural to ask what the
analogs of these Poisson algebras are in the topological--holomorphic
setting.  An answer is Poisson vertex algebras.

Another source of motivation is the presence of vertex algebras in
four-dimensional $\CN = 2$ superconformal field theories, which was
uncovered in~\cite{Beem:2013sza} and has been a subject of intensive
research for the past several years.  The classical limits of these
vertex algebras are Poisson vertex algebras.

The construction of Poisson vertex algebras we discuss in this paper
indeed provides a bridge between the two lines of developments:
whereas the Poisson vertex algebra for a four-dimensional unitary
$\CN = 2$ superconformal field theory has a canonical deformation to
the vertex operator algebra introduced in~\cite{Beem:2013sza},
dimensional reduction turns it into a Poisson algebra associated with
a TQFT in one dimension less.  We emphasize, however, that the
construction itself requires no conformal invariance and applies to a
broader class of theories.

In section~\ref{sec:PVA}, we describe the structures that define a
topological--holomorphic sector within a supersymmetric field theory,
and demonstrate that local operators in this sector comprise a Poisson
vertex algebra.  Essential for the definition of a Poisson vertex
algebra is the binary operation called the
$\lambda$-bracket~\cite{MR1651389}, which plays a role similar to the
Poisson bracket in Poisson algebras.  The $\lambda$-bracket is
constructed via a topological--holomorphic analog of topological
descent~\cite{Witten:1988ze}, much as the Poisson bracket between
local operators in a TQFT is constructed via the descent procedure.
This construction may be seen as providing a concrete physical
realization of a special instance of the Poisson additivity theorem,
proved by Nick Rozenblyum and also independently in~\cite{MR3830550}.%
\footnote{We are grateful to Dylan Butson for explaining this point to
  us.}

In section~\ref{sec:3d}, we identify topological--holomorphic sectors
in theories with $\CN = 2$ supersymmetry in three dimensions, and
determine the Poisson vertex algebras for chiral multiplets and vector
multiplets.  We content ourselves mostly with free theories here; to
go beyond that one needs to incorporate quantum corrections, both
perturbative and nonperturbative.

In section~\ref{sec:4d}, we consider Poisson vertex algebras for
$\CN = 2$ supersymmetric field theories in four dimensions.  We show
that if a theory is conformal, the associated Poisson vertex algebra
is the classical limit of a vertex algebra, which is isomorphic to the
vertex operator algebra of~\cite{Beem:2013sza} in the unitary case.
For free hypermultiplets, we confirm this relation by explicitly
computing the Poisson vertex algebra and comparing it with the vertex
operator algebra.  For gauge theories constructed from vector
multiplets and hypermultiplets, the relation to the vertex algebras
allows us to propose a description of the associated Poisson vertex
algebras.

There are many directions to explore in connection with Poisson vertex
algebras.  To conclude this introduction we briefly describe some
general ideas.

Theories related by dualities, such as mirror symmetry and
S-dualities, should have isomorphic Poisson vertex algebras.
Dualities may thus lead to interesting isomorphisms between Poisson
vertex algebras.  Conversely, isomorphisms established between Poisson
vertex algebras associated with apparently different theories may
serve as evidence that those theories are actually equivalent.

The structures considered in this paper can be further enriched by
introduction of topological--holomorphic boundaries and defects, which
themselves support Poisson vertex algebras (or vertex algebras if they
are two-dimensional and have no topological direction).  The interplay
between the Poisson vertex algebras associated with the bulk theory
and the boundaries and defects would be a fascinating topic.  In three
dimensions, relations between bulk Poisson vertex algebras and
boundary vertex algebras are discussed
in~\cite{Costello:2020ndc}.

It would also be interesting to study the Poisson vertex algebras for
theories originating from compactification of six-dimensional
$\CN = (2,0)$ superconformal field theories on Riemann
surfaces~\cite{Gaiotto:2009we,Gaiotto:2009hg} and
$3$-manifolds~\cite{Terashima:2011qi,Terashima:2011xe,Dimofte:2011ju,Dimofte:2011py}.
These Poisson vertex algebras should be geometric invariants.

Intriguing observations made in~\cite{Cordova:2015nma} suggest that
the Poisson vertex algebras for four-dimensional $\CN = 2$
supersymmetric field theories are related to wall-crossing phenomena.
Understanding this relation may be an outstanding problem.

Lastly, it would be fruitful to investigate connections to integrable
hierarchies of soliton equations.  This is the context in which the
theory of Poisson vertex algebras was originally
developed~\cite{MR2058353,MR1620526,MR2082709,MR2576030}.

\section{Poisson vertex algebras in topological--holomorphic sectors}
\label{sec:PVA}

In this paper we consider supersymmetric field theories that contain
sectors that are topological in $d \geq 1$ directions and holomorphic
in one complex direction.  The topological directions form a
$d$-dimensional manifold $M$, parametrized by real coordinates
$y = (y^i)_{i=1}^d$; the holomorphic direction is a Riemann surface
$C$ with complex coordinate $z$.  We also use $x^\mu$, $\mu = i$, $z$,
$\zb$, for the coordinates of $M \times C$, with $x^i = y^i$,
$x^z = z$ and $x^\zb = \zb$.  Typically, we take $M = \R^d$ and
$C = \C$ or a cylinder.

In this section we explain how such a topological--holomorphic sector
arises from a supersymmetric field theory, and construct a
($d$-shifted) Poisson vertex algebra in this sector.  The construction
is similar to that of the Poisson algebras for TQFTs, treated in
detail in~\cite{Beem:2018fng}.  Then, we discuss some of the basic
properties of the Poisson vertex algebra thus obtained.

\subsection{Topological--holomorphic sector}

Suppose that a quantum field theory on $M \times C$ is invariant under
translations shifting $x^\mu$, and has symmetries generated by a
fermionic conserved charge $Q$ and a fermionic one-form conserved
charge
\begin{equation}
  \Qone = \Qone_i \, \rmd y^i + \Qone_\zb \, \rmd\zb \,,
\end{equation}
satisfying the relations
\begin{align}
  \label{eq:Q2}
  Q^2 &= 0 \,,
  \\
  \label{eq:PQ}
  [Q, P_\mu] &= 0 \,
  \\
  \label{eq:PQone}
  [\Qone, P_\mu] &= 0 \,,
  \\
  \label{eq:QQone}
  [Q, \Qone] &= \iu P_i \, \rmd y^i + \iu P_\zb \, \rmd\zb \,.
\end{align}
%
% We don't need  $\Qone \wedge \Qone = 0$.
%
Here $P_\mu$ is the generator of translations in the
$x^\mu$-direction, acting on a local operator $\CO$ by%
\footnote{In the path integral formulation, the action of a conserved
  charge $X$ on a local operator $\CO$ located at a point $x$ is
  implemented by the integration of the Hodge dual of the associated
  conserved current over a simply-connected codimension-$1$ cycle
  surrounding $x$.  If we take this cycle to be the difference of two
  time slices sandwiching $x$, we obtain the familiar formula
  $X \cdot \CO = [X, \CO]$ as an operator acting on the Hilbert space
  of states.  The cycle can also be infinitesimally small, so the
  operation $\CO \mapsto X \cdot \CO$ is local.}
\begin{equation}
  \label{eq:PmuO}
  \iu P_\mu \cdot \CO = \del_\mu \CO \,.
\end{equation}
The graded commutator $[\ , \ ]$ is defined by
$[a,b] = ab - (-1)^{F(a) F(b)} ba$, where $(-1)^{F(a)} = +1$ if $a$ is
bosonic and $-1$ if fermionic.

Since $Q$ squares to zero, we can use it to define a cohomology in the
space of operators.  The commutation relations~\eqref{eq:PQ}
and~\eqref{eq:QQone} show that $P_\mu$ act on the $Q$-cohomology, and
$P_i$ and $P_\zb$ act trivially.  Therefore, the $Q$-cohomology class
$\ec{\CO(z)}$ of a $Q$-closed local operator $\CO(y, z, \zb)$ is
independent of its position in $M$ and varies holomorphically on $C$.
The same is true for correlation functions of $Q$-closed local
operators, which depend only on the $Q$-cohomology classes and not the
choices of representatives.

Given multiple $Q$-closed local operators $\CO_a(x_a)$, $a = 1$,
$\dotsc$, $n$, located at $x_a = (y_a, z_a, \zb_a)$, we define the
product of their $Q$-cohomology classes $\ec{\CO_a(z_a)}$ by
\begin{equation}
  \ec{\CO_1(z_1)} \dotsm \ec{\CO_n(z_n)}
  = \ec{\CO_1(x_1) \dotsm \CO_n(x_n)} \,,
  \quad
  \text{$x_a \neq x_b$ if $a \neq b$} \,,
\end{equation}
displacing the operators in $M$ if necessary.  (Note that this
definition requires $d \geq 1$.)  The condition that no two operators
are located at the same point in $M \times C$ is important because a
product of coincident operators is generally singular in a quantum
field theory.

As long as no two operators are located at the same point in $M$, we
may set all $z_a$ equal in the above definition of product.  Pointwise
multiplication on $C$ furnishes the $Q$-cohomology of local operators
with the structure of an algebra, which is furthermore equipped with a
derivation $P_z$.  We denote this algebra by $\CV$.

While the space of operators is always $\Z_2$-graded by the fermion
parity $(-1)^F$, in many examples this grading is refined by a
$\Z$-grading by a weight~$F$ of a $\U(1)$ symmetry such that
\begin{equation}
  F(Q) = -F(\Qone) = 1 \,,
  \qquad
  F(P_\mu) = 0 \,.
\end{equation}
In that case, $\CV$ is $\Z$-graded by $F$.  It may also happen that
only a $\Z_{2n}$ subgroup of such a $\U(1)$ symmetry is unbroken and
$\CV$ is $\Z_{2n}$-graded.  There may be additional $\U(1)$ symmetries
providing further gradings.

We will refer to the collection of various structures associated with
the $Q$-cohomology described above as the
\emph{topological--holomorphic sector} of the theory.  This definition
is somewhat weaker than the usual notion of a topological--holomorphic
field theory of cohomological type, in which one requires that the
components $T_{ij}$ and $T_{\mu\zb} = T_{\zb\mu}$ of the
stress--energy tensor are $Q$-exact.

\subsection{Topological--holomorphic descent}

In cohomological TQFTs, there is a procedure called topological
descent~\cite{Witten:1988ze} which produces $Q$-cohomology classes of
nonlocal operators starting from a $Q$-cohomology class of local
operator.  This construction can be adapted in the
topological--holomorphic setting.

For a local operator $\CO$, let us define its $k$th descendant
$\CO^{(k)}$ by
\begin{equation}
  \CO^{(k)}
  =
  \frac{1}{k!}
  \rmd x^{\mu_1} \wedge \dotsb \wedge \rmd x^{\mu_k}
  (\Qone_{\mu_1} \dotsm \Qone_{\mu_k}) \cdot \CO \,.
\end{equation}
Here we have set $\Qone_z = 0$.  We denote the total descendent of
$\CO$ by $\CO^*$:
\begin{equation}
  \CO^*
  = \sum_{k=0}^{d+1} \CO^{(k)}
  = \exp(\Qone) \cdot \CO \,.
\end{equation}
Similarly, we introduce
\begin{equation}
  \CO^{-*} = \exp(-\Qone) \cdot \CO \,.
\end{equation}
These descendant operators satisfy
\begin{equation}
  \label{eq:QO*}
  Q \cdot \CO^{\pm*}
  = \pm\rmd' \CO^{\pm*} + (Q \cdot \CO)^{\mp*} \,,
\end{equation}
where $\CO^{+*} = \CO^*$ and
\begin{equation}
  \rmd' = \rmd y^i \del_i + \rmd\zb \del_\zb \,.
\end{equation}

Now suppose that $\CO$ is a $Q$-closed local operator.  Then, for any
holomorphic one-form $\omega(z) = \omega_z(z) \, \rmd z$ on $C$, we
have
\begin{equation}
  \label{eq:QomegaO*}
  Q \cdot (\omega \wedge \CO^*)
  = -\rmd(\omega \wedge \CO^*) \,.
\end{equation}
Therefore, for any cycle $\Gamma \subset M \times C$, the integral
\begin{equation}
  \label{eq:intomegaO*}
  \int_\Gamma \omega \wedge \CO^*
\end{equation}
is a $Q$-closed operator.  The $Q$-cohomology class of this operator
depends on $\Gamma$ only through the homology class
$\ec{\Gamma} \in H_\bullet(M \times C)$ since
$\rmd(\omega \wedge \CO^*) = 0$ in the $Q$-cohomology.  It also
depends on $\CO$ only through the $Q$-cohomology class $\ec{\CO}$
because shifting $\CO$ by $Q \cdot \COt$ just changes the integral by
$Q \cdot \int_\Gamma \omega \wedge \COt^{-*}$.  We call this procedure
of constructing nonlocal $Q$-cohomology classes from local ones
\emph{topological--holomorphic descent}.

More generally, for a product $\CO_1(x_1) \dotsm \CO_n(x_n)$ of $n$
$Q$-closed local operators, subject to the condition that
$x_a \neq x_b$ if $a \neq b$, we define its descendant as a
differential form on the configuration space of $n$ points on
$M \times C$,
\begin{equation}
  \conf_n(M \times C)
  =
  \{(x_1, \dotsc, x_n) \in (M \times C)^n
  \mid \text{$x_a \neq x_b$ if $a \neq b$}\} \,.
\end{equation}
Let $\pi_a\colon \conf_n(M \times C) \to M \times C$ be the projection
to the $a$th factor, and for differential forms $\alpha_1$, $\dotsb$,
$\alpha_n \in \Omega^\bullet(M \times C)$ let
\begin{equation}
  \alpha_1 \boxtimes \dotsb \boxtimes \alpha_n
  =
  \pi_1^*\alpha_1 \wedge \dotsb \wedge \pi_n^*\alpha_n
  \in \Omega^\bullet(\conf_n(M \times C)) \,.
\end{equation}
In this language, we may think of the product of local operators at
$n$ distinct points as the evaluation of a local operator at a single
point in $\conf_n(M \times C)$:
\begin{equation}
  \CO_1(x_1) \dotsm \CO_n(x_n)
  = (\CO_1 \boxtimes \dotsb \boxtimes \CO_n)(x_1, \dotsc, x_n) \,.
\end{equation}

We define the descendant of $\CO_1 \boxtimes \dotsb \boxtimes \CO_n$
to be
\begin{equation}
  \label{eq:(O1...On)*}
  (\CO_1 \boxtimes \dotsb \boxtimes \CO_n)^*
  =
  \CO_1^* \boxtimes \sigma^{F_1} \CO_2^*
  \boxtimes \dotsb \boxtimes \sigma^{F_1 + \dotsb + F_{n-1}} \CO_n^* \,,
\end{equation}
where $F_a = F(\CO_a)$ and the operator $\sigma$ multiplies even forms
by $+1$ and odd forms by $-1$.  The descendant satisfies
\begin{equation}
  Q \cdot (\CO_1 \boxtimes \dotsb \boxtimes \CO_n)^*
  = \rmd'(\CO_1 \boxtimes \dotsb \boxtimes \CO_n)^* \,,
\end{equation}
with $\rmd'$ acting on $\Omega^\bullet(\conf_n(M \times C))$ in the
obvious manner.  The role of $\sigma$ in the
definition~\eqref{eq:(O1...On)*} is to supply, via the relation
$\sigma \rmd' = -\rmd' \sigma$, a factor of
$(-1)^{F_1 + \dotsb + F_{a-1}}$ when $\rmd'$ acts on $\CO_a^*$.

Given a holomorphic top-form $\omega$ on $C^n$ (or more precisely, its
pullback to $\conf_n(M \times C)$) and a cycle
$\Gamma \subset \conf_n(M \times C)$, the integral
\begin{equation}
  \int_\Gamma \omega \wedge (\CO_1 \boxtimes \dotsb \boxtimes \CO_n)^*
\end{equation}
is a $Q$-closed operator whose $Q$-cohomology class depends only on
$\omega$, the homology class $\ec{\Gamma}$ and the $Q$-cohomology classes
$\ec{\CO_a}$.

\subsection{Secondary products}

The algebra $\CV$ of $Q$-cohomology of local operators possesses
\emph{secondary products}, in addition to the ordinary (or primary)
product given by pointwise multiplication on $C$.  These products make
use of topological--holomorphic descent.

Choose a translation invariant holomorphic two-form
$\kappa(z_1,z_2) = \kappa_{z_1 z_2}(z_1,z_2) \rmd z_1 \wedge \rmd z_2$
on $C \times C$.  By translation invariance we mean
\begin{equation}
  \label{eq:delomega}
  \CL_{\del_{z_1} + \del_{z_2}} \kappa = 0 \,,
\end{equation}
or $(\del_{z_1} + \del_{z_2}) \kappa_{z_1 z_2}(z_1,z_2) = 0$, where
$\CL_V$ denotes the Lie derivative with respect to $V$.  For two
$Q$-closed local operators $\CO_1$ and $\CO_2$, we define a new
$Q$-closed operator $\CO_1 \star \CO_2$ by
\begin{equation}
  (\CO_1 \star_\kappa \CO_2)(x_2)
  =
  \biggl(\int_{S^{d+1}_{x_2}} \iota_{\del_{z_2}} \kappa(z_1,z_2)
  \wedge \CO_1^{(d)}(x_1)\biggr)
  \CO_2(x_2) \,,
\end{equation}
where $S^{d+1}_{x_2}$ is a $(d+1)$-sphere centered at the point $x_2$
at which $\CO_2$ is placed, with the radius taken to be sufficiently
small.  On this operator $\iu P_\mu$ acts as $\del_\mu$ since
infinitesimally displacing $\CO_1$ in the integrand is equivalent to
shifting the center of $S^{d+1}_{x_2}$, which does not change the
homology class, plus shifting the argument $z_2$ in
$\iota_{\del_{z_2}} \kappa(z_1,z_2)$.  In particular, the
relation~\eqref{eq:delomega} is crucial for the
topological--holomorphic descent to hold for
$\CO = \CO_1 \star_\kappa \CO_2$.

The $Q$-cohomology class $\ec{\CO_1 \star_\kappa \CO_2}$ depends only
on $\ec{\CO_1}$ and $\ec{\CO_2}$.  Hence, $\star_\kappa$ defines a
secondary product for $\CV$.  It has cohomological degree $-d$ since
$\CO_1 \star_\kappa \CO_2$ uses the $d$th descendant of $\CO_1$
which has fermion number $F_1 - d$.

The secondary product $\star_\kappa$ acts on the primary product as a
derivation: it satisfies the \emph{Leibniz rule}
\begin{equation}
  \label{eq:Leibniz}
  \ec{\CO_1} \star_\kappa (\ec{\CO_2} \ec{\CO_3})
  =
  (\ec{\CO_1} \star_\kappa \ec{\CO_2}) \ec{\CO_3}
  + (-1)^{(F_1 + d) F_2}
    \ec{\CO_2} (\ec{\CO_1} \star_\kappa \ec{\CO_3}) \,.
\end{equation}
This identity simply says that $S^{d+1}$ enclosing both $\CO_2$ and
$\CO_3$ can be divided into two $S^{d+1}$, one containing $\CO_2$ and
the other containing $\CO_3$.  Such a decomposition is possible
because $\CO_2$ and $\CO_3$ in the product $\ec{\CO_2} \ec{\CO_3}$ are by
definition separated in $M$.

\subsection{\texorpdfstring{$\lambda$-bracket}{lambda-bracket}}

Secondary products are defined locally; we can make $S^{d+1}_{x_2}$ as
small as we wish, for changing the radius does not affect its homology
class.  Then, by the locality of quantum field theory (expressed, for
example, as the gluing axiom in the Atiyah--Segal formulation), any
secondary product can be computed purely based on the knowledge of the
behavior of the theory in a neighborhood
$U \times V \subset M \times C$ of the point at which we are taking
the product, where $U \times V \simeq \R^d \times \C$ topologically.

On $V$, we can expand $\kappa(z_1,z_2)$ around $z_2$:
\begin{equation}
  \kappa(z_1,z_2)
  =
  \sum_{k=0}^\infty \frac{1}{k!}
  (z_1 - z_2)^k \del_{z_1}^k \kappa_{z_1 z_2}(z_2,z_2)
  \, \rmd z_1 \wedge \rmd z_2 \,.
\end{equation}
Since $\star_\kappa$ is linear in $\kappa$, all information about
secondary products in the neighborhood $U \times V$ is encoded in the
cases when $\kappa(z_1,z_2) = (z_1 - z_2)^k \rmd z_1 \wedge \rmd z_2$.
The generating function of such forms is
$e^{\lambda(z_1 - z_2)} \rmd z_1 \wedge \rmd z_2$, where $\lambda$ is
a formal variable.

This consideration motivates us to introduce the
\emph{$\lambda$-bracket}~\cite{MR1651389}:
\begin{equation}
  \label{lbracket}
  \lbracket{\lambda}{\ec{\CO_1}}{\ec{\CO_2}}(z_2)
  =
  (-1)^{F_1 d} \biggec{\biggl(\int_{S^{d+1}_{x_2}}
  e^{\lambda(z_1 - z_2)} \rmd z_1 \wedge \CO_1^{(d)}(x_1) \biggr)
  \CO_2(x_2)} \,.
\end{equation}
The spacetime is here taken to have the topology of $\R^d \times \C$,
but not necessarily given the standard metric.  In general, the
$\lambda$-bracket may depend on the choice of metric and other
geometric structures.

The $\lambda$-bracket satisfies a few important identities, besides
the Leibniz rule~\eqref{eq:Leibniz}.  First, it has
\emph{sesquilinearity}:
\begin{equation}
  \label{eq:sesquilinearity}
  \begin{aligned}
    \lbracket{\lambda}{\del_z \ec{\CO_1}}{\ec{\CO_2}}
    &=
    -\lambda \lbracket{\lambda}{\ec{\CO_1}}{\ec{\CO_2}} \,,
    \\
    \lbracket{\lambda}{\ec{\CO_1}}{\del_z \ec{\CO_2}}
    &=
    (\lambda + \del_z) \lbracket{\lambda}{\ec{\CO_1}}{\ec{\CO_2}} \,.    
  \end{aligned}
\end{equation}
Second, it has the following \emph{symmetry} under exchange of the two
arguments:
\begin{equation}
  \label{eq:symmetry}
  \lbracket{\lambda}{\ec{\CO_1}}{\ec{\CO_2}}
  =
  -(-1)^{(F_1 + d)(F_2 + d)}
  \lbracket{-\lambda - \del_z}{\ec{\CO_2}}{\ec{\CO_1}} \,.
\end{equation}
The $\lambda$-bracket with $\lambda$ being an operator $\Lambda$ is to
be understood as a power series in $\Lambda$ acting on the local
operators in the bracket:
\begin{equation}
  \lbracket{\Lambda}{\ec{\CO_2}}{\ec{\CO_1}}
  =
  \bigl(\exp(\Lambda \del_\lambda)
  \lbracket{\lambda}{\ec{\CO_2}}{\ec{\CO_1}}\bigr)\bigr|_{\lambda=0} \,.
\end{equation}
Finally, the $\lambda$-bracket satisfies the \emph{Jacobi identity}
\begin{multline}
  \label{eq:Jacobi}
  \lbracket{\lambda}{\ec{\CO_1}}{\lbracket{\mu}{\ec{\CO_2}}{\ec{\CO_3}}}
  \\
  =
  \lbracket{\lambda + \mu}{\lbracket{\lambda}{\ec{\CO_1}}
    {\ec{\CO_2}}}{\ec{\CO_3}}
  +
  (-1)^{(F_1+ d)(F_2+d)}
  \lbracket{\mu}{\ec{\CO_2}}
  {\lbracket{\lambda}{\ec{\CO_1}}{\ec{\CO_3}}} \,.
\end{multline}

Let us prove the above identities one by one.  Since these are
equalities between holomorphic functions on $\C$, it suffices to
demonstrate that they hold on $\C^\times = \C \setminus \{0\}$.  Thus
we replace the spacetime with $\R^d \times \C^\times$ during the
derivation.

The sesquilinearity~\eqref{eq:sesquilinearity} is straightforward to
show.  The first equation follows from the fact that to the integral
appearing in the definition~\eqref{lbracket} of the $\lambda$-bracket,
an infinitesimal displacement of $\CO_1$ in the $z$-direction has the
same effect as shifting the parameter $z_2$ in $S^{d+1}_{x_2}$ and
$e^{\lambda(z_1 - z_2)}$.  To prove the second equation, one may just
take a derivative of the $\lambda$-bracket with respect to $z_2$.

To establish the symmetry~\eqref{eq:symmetry}, we show
\begin{multline}
  \int_{S^1_0} f(z_2) \, \rmd z_2 \lbracket{\lambda}{\ec{\CO_1}}{\ec{\CO_2}}(z_2)
  \\
  =
  -(-1)^{(F_1 + d)(F_2 + d)}
  \int_{S^1_0} f(z_1) \, \rmd z_1
  \lbracket{-\lambda - \del_z}{\ec{\CO_2}}{\ec{\CO_1}}(z_1)
\end{multline}
for any holomorphic function $f$ on $\C^\times$, where $S^1_0$ is a
circle around $z = 0$.  To this end we adopt the point of view of the
configuration space.  As an integral in
$\conf_2(\R^d \times \C^\times)$, the left-hand side is given by
\begin{equation}
  (-1)^{F_1 d} \biggec{\int_{S^{d+1}_{x_2} \times S^1_0}
  f(z_2) \, \rmd z_2 \wedge e^{\lambda z_{12}} \rmd z_{12}
  \wedge (\CO_1 \boxtimes \CO_2)^*} \,.
\end{equation}
Here we have defined $x_{12}^\mu = x_1^\mu - x_2^\mu$ and think of
$(x_{12}^\mu, x_2^\mu)$ as coordinates on
$\conf_2(\R^d \times \C^\times)$.  The cycle
$S^{d+1}_{x_2} \times S^1_0 \subset \conf_2(\R^d \times \C^\times)$ is
homologous to $(-1)^d S^1_0 \times S^{d+1}_{x_1}$,%
\footnote{Since $H_{d+2}(\conf_2(\R^d \times \C^\times)) \iso \Z$, to
  establish this relation we just need to evaluate a suitable
  cohomology class on these cycles.  We can take this class to be
  $\rmd(\theta_1 + \theta_2) \wedge \rmd^{-1}(\bigwedge_\mu
  \delta(x_{12}^\mu) \rmd x_{12}^\mu)$, where $\theta = \arg z$ and
  $\delta(x)$ is the delta function.  To compute the homology group,
  let $U = \conf_2(\R^d \times \C^\times)$ and
  $V \simeq \R^{2d+3} \times S^1$ be a normal neighborhood of the
  diagonal $\Delta$ of $(\R^d \times \C^\times)^2$.  Then,
  $U \cup V = (\R^d \times \C^\times)^2 \simeq \R^{2d+2} \times S^1
  \times S^1$ and
  $U \cap V = V \setminus \Delta \simeq \R^{d+2} \times S^1 \times
  S^{d+1}$.  From the Mayer--Vietoris sequence
  \begin{equation}
    \dotsb \to H_{d+3}(U \cup V) \to H_{d+2}(U \cap V)
    \to H_{d+2}(U) \oplus H_{d+2}(V)
    \to H_{d+2}(U \cup V) \to \dotsb \,,
  \end{equation}
  we get $H_{d+2}(U) \iso H_{d+2}(S^1 \times S^{d+1}) \iso \Z$.}
so we can rewrite
this integral as
\begin{equation}
  -(-1)^{F_1 d + d}
  \biggec{\int_{S^1_0 \times S^{d+1}_{x_1}} f(z_2) \rmd z_1 \wedge
    e^{-\lambda z_{21}} \rmd z_{21} \wedge (\CO_1 \boxtimes \CO_2)^*} \,.
\end{equation}

We expand $f(z_2)$ around $z_1$ and replace the powers of $z_{21}$
in the series with powers of $-\del_\lambda$ to get
\begin{multline}
  -(-1)^{F_1 d + d} \sum_{k=0}^\infty \frac{1}{k!}
  (-\del_\lambda)^k
  \biggec{\int_{S^1_0 \times S^{d+1}_{x_1}} \del_z^k f(z_1) \rmd z_1 \wedge
    e^{-\lambda z_{21}} \rmd z_{21} \wedge (\CO_1 \boxtimes \CO_2)^*}
  \\
  =
  -(-1)^{F_1 d + d} \sum_{k=0}^\infty \frac{1}{k!}
  (\iu P_z\del_\lambda)^k
  \biggec{\int_{S^1_0 \times S^{d+1}_{x_1}} f(z_1) \rmd z_1 \wedge
    e^{-\lambda z_{21}} \rmd z_{21} \wedge (\CO_1 \boxtimes \CO_2)^*}
  \,,
\end{multline}
In the equality we have used the invariance of the $Q$-cohomology
class of the integral under a shift of the cycle in the $z$-direction.
The last expression is equal to
\begin{equation}
  -(-1)^{F_1 d + d + F_1 d + F_1 (F_2 + d) + F_2 d}
  \int_{S^1_0} f(z_1) \, \rmd z_1
  \exp(\iu P_z \del_\lambda)
  \lbracket{-\lambda}{\ec{\CO_2}}{\ec{\CO_1}}(z_1) \,,
\end{equation}
where a factor of $(-1)^{F_1 d}$ comes from $\sigma^{F_1}$ in the
definition~\eqref{eq:(O1...On)*} of $(\CO_1 \boxtimes \CO_2)^*$, the
factor $(-1)^{F_1 (F_2 + d)}$ is due to an interchange of
$\CO_1^{(0)}$ and $\CO_2^{(d)}$, and a factor of $(-1)^{F_2 d}$ comes
from $\sigma^{F_2}$ in $(\CO_2 \boxtimes \CO_1)^*$.  This is what we
wanted.

The Jacobi identity~\eqref{eq:Jacobi}, like the Leibniz rule, is based
on the decomposition of a large $S^{d+1}$ surrounding both $\CO_2$ and
$\CO_3$ into a small $S^{d+1}$ around $\CO_2$ and another small
$S^{d+1}$ around $\CO_3$.  The latter small $S^{d+1}$ gives the second
term on the right-hand side.

Obtaining the first term is a little tricky.  The shift in $\mu$ is
easy to understand.  What is tricky is that this term involves
$(\CO_1^* \CO_2)^*$, while the left-hand side only contains
$\CO_1^* \CO_2^*$, which differs by terms involving $\Qone_\mu$ acting
on $\CO_1^*$.  To avoid getting these extra terms, we can use the
symmetry~\eqref{eq:symmetry}.

Let us switch to the configuration space viewpoint, as we did in the
proof of the symmetry.  The relevant cycle in
$\conf_3(\R^d \times \C^\times)$ is one that represents the situation
in which $\CO_1^*$ is integrated over $S^{d+1}$ containing $\CO_2^*$,
while $\CO_2^*$ is integrated over $S^{d+1}$ around $\CO_3^*$, and
finally $\CO_3^*$ is integrated over $S^1$ around the origin of
$\C^\times$.  Up to a sign, this cycle is homologous to one in which
$\CO_2^*$ is surrounded by $\CO_1^*$, and by $\CO_3^*$ farther back,
and itself circles around the origin of $\C^\times$.%
\footnote{Think of $\CO_1^*$, $\CO_2^*$ and $\CO_3^*$ as the moon, the
  earth and the sun, respectively, and consider the heliocentric
  versus geocentric descriptions of their relative motion.  The origin
  of $\C^\times$ can be the center of our galaxy.}
Again up to a sign, the second cycle gives
$\lbracket{-\lambda - \mu -
  \del_z}{\ec{\CO_3}}{\lbracket{\lambda}{\ec{\CO_1}}{\ec{\CO_2}}}$ and
equals the term in question in the Jacobi identity.  To determine the
sign, we can go back to the original viewpoint and consider the
special case when $\Qone_\mu$ annihilate
$\int_{S^{d+1}_{x_2}} e^{\lambda(z_1 - z_2)} \rmd z_1 \wedge \CO_1^*$.

\subsection{Poisson vertex algebra}

The $Q$-cohomology of local operators $\CV$ is a graded
$\C[P_z]$-module endowed with the $\lambda$-bracket
$\lbracket{\lambda}{\ }{\ }\colon \CV \otimes \CV \to \CV[\lambda]$
with fermion number
\begin{equation}
  F(\lbracket{\lambda}{\ }{\ }) = - d \,,
\end{equation}
satisfying the sesquilinearity~\eqref{eq:sesquilinearity}, the
symmetry~\eqref{eq:symmetry} and the Jacobi
identity~\eqref{eq:Jacobi}.  These properties make $\CV$ a
\emph{$d$-shifted Lie conformal algebra}~\cite{MR1651389}.

In addition, the $\lambda$-bracket satisfies the Leibniz
rule~\eqref{eq:Leibniz}.  A unital commutative associative $d$-shifted
Lie conformal algebra, with the Leibniz rule obeyed, is called a
\emph{$d$-shifted Poisson vertex algebra}.  Thus, we have shown that
the algebra of local operators in the topological--holomorphic sector
of a quantum field theory on $M \times C$ possesses the structure of a
$d$-shifted Poisson vertex algebra.  We refer the reader to
\cite{MR3751122} for an introduction to Poisson vertex algebras.

Suppose that the theory has rotation symmetry on $C$, and let $J$ be
its generator.  We normalize $J$ in such a way that $P_z$ has $J = 1$.
If $Q$ has a definite spin $J(Q)$ (as opposed to being a linear
combination of operators with different spins), then $\CV$, as an
algebra, is also graded by $J$.  In this case we can choose $\Qone$ to
have spin
\begin{equation}
  J(\Qone) = -J(Q) \,.
\end{equation}

As it is, the $\lambda$-bracket is not compatible with this grading.
The problem is that for computation of the $\lambda$-bracket, one of
the local operators needs to be displaced from the center of rotations
$x_2$, and while $\ec{S^{d+1}_{x_2}}$ used in the $\lambda$-bracket is
invariant under rotations, the factor $e^{\lambda(z_1 - z_2)}$ is not.
This means that the $\lambda$-bracket, when expanded in $\lambda$,
consists of infinitely many parts carrying different values of $J$.
We can, however, remedy this problem if we simultaneously shift the
phase of $\lambda$.  In other words, the $\lambda$-bracket respects
the $J$-grading if we assign
\begin{equation}
  J(\lambda) = 1 \,.
\end{equation}
Since its definition contains the one-form $\rmd z_1$ and $d$ copies
of $\Qone$, the $\lambda$-bracket changes the $J$-grading by
\begin{equation}
  J(\lbracket{\lambda}{\ }{\ }) = -J(Q) d - 1 \,.
\end{equation}

\subsection{Dimensional reduction to the Poisson algebra of a TQFT}

Poisson vertex algebras are generalizations of Poisson algebras, which
are commutative associative algebras endowed with a Lie bracket
obeying the Leibniz rule.  Indeed, the quotient of $\CV$ by the ideal
$(\del_z\CV) \CV$ is a $d$-shifted Poisson algebra with respect to
the Poisson bracket induced from the $\lambda^0$-order part
$\lbracket{\lambda}{\ }{\ }|_{\lambda = 0}$ of the $\lambda$-bracket.

This Poisson algebra may be interpreted as a structure associated with
the $(d+1)$-dimensional theory that one obtains by taking $C$ to be a
cylinder and performing dimensional reduction on the circumferential
direction.  Since local $Q$-cohomology classes vary holomorphically on
$C$, after the reduction they become independent of their positions in
the longitudinal direction.  Thus, the topological--holomorphic sector
on $M \times C$ is turned into a topological sector on $M \times \R$.
The latter carries a Poisson bracket of degree $-d$ on the algebra of
local operators~\cite{Beem:2018fng}, and in favorable situations,
$\CV/(\del_z\CV) \CV$ coincides with this Poisson algebra.

In general, the relation between the two Poisson algebras can be more
complicated.  For instance, the theory on $M \times \R$ may have local
operators that come from line operators in $M \times C$ extending in
the reduced direction.  If these local operators are $Q$-closed, they
may represent $Q$-cohomology classes that are not present in
$\CV/(\del_z\CV)\CV$.  If they are not $Q$-closed, then some elements
of $\CV/(\del_z\CV)\CV$ may be paired up with them and get annihilated
from the $Q$-cohomology.

\subsection{Lie algebra of zero modes}

As a vector space, we can also consider the quotient
$\CVb = \CV/\del_z \CV$.  This is the space of zero modes of local
$Q$-cohomology classes.  The quotient map $\CV \to \CVb$ is denoted by
$\zmode$.

The space $\CVb$ is actually a $d$-shifted Lie algebra, with the Lie
bracket $\{\ ,\ \}$ of degree $-d$ given by
\begin{equation}
  \{\zmode\ec{\CO_1}, \zmode\ec{\CO_2}\}
  = \zmode \lbracket{\lambda}{\ec{\CO_1}}{\ec{\CO_2}}|_{\lambda = 0} \,.
\end{equation}
Moreover, the Lie algebra $\CVb$ acts on $\CV$ by
\begin{equation}
  \{\zmode\ec{\CO_1}, \ec{\CO_2}\}
  = \lbracket{\lambda}{\ec{\CO_1}}{\ec{\CO_2}}|_{\lambda = 0} \,.
\end{equation}
As such, $\CVb$ may be regarded as a Lie algebra generating a
continuous symmetry of $\CV$.  A proof of these statements is
straightforward~\cite{MR3751122}.

\section{\texorpdfstring{$\CN = 2$ supersymmetric field theories in three
    dimensions}{N = 2 supersymmetric field theories in three
  dimensions}}
\label{sec:3d}

The lowest dimensionality in which the structures of
topological--holomorphic sectors arise is three.  In this section, we
define Poisson vertex algebras for three-dimensional $\CN = 2$
supersymmetric field theories and determine them in basic examples.
We will take
\begin{equation}
  M \times C = \R \times \C
\end{equation}
and denote the coordinate on $\R$ by $t$.

\subsection{Topological--holomorphic sector}

The $\CN = 2$ supersymmetry algebra in $2+1$ dimensions has four
supercharges $Q_\alpha$, $\Qb_\alpha$, $\alpha = \pm$, satisfying the
commutation relations%
\footnote{The $\CN = 2$ supersymmetry algebra in $2+1$ dimensions can
  be obtained from the $\CN = 1$ supersymmetry algebra in $3+1$
  dimensions by dimensional reduction.  For the latter algebra we
  follow the conventions of~\cite{Wess:1992cp}, except that we rescale
  the supercharges by a factor of $1/\sqrt{2}$.  We have chosen to
  perform the reduction along the $x^2$-direction and subsequently
  renamed $x^3$ to $x^2$.}
\begin{align}
  \label{eq:3dN=2SUSY-1}
  [Q_\pm, \Qb_\pm]
  &= -P_0 \pm P_2 \,,
  \\
  \label{eq:3dN=2SUSY-2}
  [Q_\pm, \Qb_\mp]
  &= P_1 \mp \iu Z \,,
  \\
  \label{eq:3dN=2SUSY-3}
  [Q_\alpha, Q_\beta]
  &=
  [\Qb_\alpha, \Qb_\beta]
  =
  0 \,,
\end{align}
where $Z$ is a real central charge.  Performing the Wick rotation
$x^0 \to -\iu x^3$ and introducing the complex coordinate
$z = (x^2 + \iu x^3)/2$, the commutation
relations~\eqref{eq:3dN=2SUSY-1} become
\begin{equation}
  [Q_+, \Qb_+]
  = P_z \,,
  \qquad
  [Q_-, \Qb_-]
  = -P_\zb \,.
\end{equation}

Taking the $x^1$-direction as the topological direction $M = \R$ and
the $z$-plane as the holomorphic direction $C = \C$, we wish to find a
supercharge $Q$ such that $Q^2 = 0$ and $P_t$, $P_\zb$ are $Q$-exact.
Moreover, $P_z$ should not appear in any commutators between $Q$ and
other generators so that it is unconstrained in the $Q$-cohomology.
These conditions require $Q$ to take the form
\begin{equation}
  Q = a Q_- + d \Qb_- \,,
\end{equation}
with $a$, $d$ being complex numbers.  We have $Q^2 = -ad P_\zb$, so we
must set either $a$ or $d$ to zero.  Without loss of generality, we
can take
\begin{equation}
  Q = \Qb_- \,.
\end{equation}

From the commutation
relations~\eqref{eq:3dN=2SUSY-1}--\eqref{eq:3dN=2SUSY-3}, we see that
in general $P_t$ cannot be genuinely $Q$-exact; rather, the
combination $P_t - \iu Z$ is $Q$-exact.  An obvious way to construct a
Poisson vertex algebra in such a situation is to take the
$Q$-cohomology in the subalgebra of local operators that have $Z = 0$.

Instead, we may simply regard $P_t - \iu Z$ as generating a modified
translation symmetry in the $t$-direction, which is ``twisted'' by the
central charge.  Given a local operator $\CO$, we define its
twisted-translated counterpart ${}^Z\CO$ by
\begin{equation}
  {}^Z\CO(t,z,\zb)
  = \exp\bigl(\iu(P_t - \iu Z)t\bigr) \cdot \CO(0, z,\zb)
  = \exp(Zt) \cdot \CO(t, z,\zb) \,.
\end{equation}
Then, the $Q$-exact operator $\iu(P_t - \iu Z)$ acts on ${}^Z\CO$ as
$\del_t$.  We can thus construct a Poisson vertex algebra from local
operators translated by $P_t - \iu Z$.

In many cases, there is a $\U(1)$ R-symmetry under which $Q_\alpha$ has
charge $R = -1$ and $\Qb_\alpha$ has $R = 1$.  Requiring
$R(Q) = - R(\Qone) = 1$ fixes the one-form supercharge:
\begin{equation}
  \Qone = \iu Q_+ \rmd t - \iu Q_- \rmd\zb \,.
\end{equation}
We have $J(Q) = -J(\Qone) = -1/2$, so the $\lambda$-bracket has
$(R,J) = (-1,-1/2)$.

It is often useful to twist the theory by regarding
\begin{equation}
  J' = J + \frac{R}{2}
\end{equation}
as a generator of rotations on $C$.  Under the twisted rotations,
$J'(Q) = J'(\Qone) = 0$, which means that $Q$ is a scalar and $\Qone$
transforms correctly as a differential form.  We have
\begin{equation}
  J'(\lbracket{\lambda}{\ }{\ }) = -1 \,.
\end{equation}

\subsection{Free chiral multiplets}
\label{sec:3d-free-chiral}

Let us determine the Poisson vertex algebra for a free theory of
chiral multiplets.  This is arguably the simplest $\CN = 2$
supersymmetric field theory.

A chiral multiplet $\Phi$ consists of a complex bosonic scalar $\phi$,
a pair of fermionic spinors $\psi_\pm$ and a complex bosonic scalar
$F$.  Under the action of the linear combination
$\eps_- Q_+ - \eps_+ Q_- - \epsb_- \Qb_+ + \epsb_+ \Qb_-$ of the
supercharges, these fields transform as
% \begin{equation}
%   \begin{aligned}
%     \delta\phi
%     &= \eps_-\psi_+ - \eps_+\psi_- \,,
%     \\
%     \delta\psi_+
%     &= \iu\epsb_- D_z\phi - \iu\epsb_+ D_w\phi + \eps_+ F \,,
%     \\
%     \delta\psi_-
%     &= \iu\epsb_- D_\wb\phi + \iu\epsb_+ D_\zb\phi + \eps_- F \,,
%     \\
%     \delta F
%     &= -\iu\epsb_+(D_\zb\psi_+ + D_w\psi_- - \iu\sqrt2\lambdab_-)
%        - \iu\epsb_-(D_\wb\psi_+ - D_z\psi_- + \iu\sqrt2\lambdab_+) \,,
%   \end{aligned}
% \end{equation}
% \begin{equation}
%   D_w = D_1 + \sigma \,,
%   \qquad
%   D_\wb = D_1 - \sigma \,,
% \end{equation}
\begin{equation}
  \begin{aligned}
    \delta\phi
    &= \eps_-\psi_+ - \eps_+\psi_- \,,
    \\
    \delta\psi_+
    &= \iu\epsb_- \del_z\phi - \iu\epsb_+ (\del_t + m)\phi + \eps_+ F \,,
    \\
    \delta\psi_-
    &= \iu\epsb_- (\del_t - m)\phi + \iu\epsb_+ \del_\zb\phi + \eps_- F \,,
    \\
    \delta F
    &= -\iu\epsb_+\bigl(\del_\zb\psi_+ + (\del_t + m)\psi_-\bigr)
       - \iu\epsb_-\bigl((\del_t - m)\psi_+ - \del_z\psi_-\bigr) \,,
  \end{aligned}
\end{equation}
where $m$ is a real mass parameter called the twisted mass.  The
supersymmetry transformations for the conjugate antichiral multiplet
$\Phib = (\phib, \psib_\pm, \Fb)$ can be obtained by hermitian
conjugation, which exchanges unbarred fields to the corresponding
barred fields and vice versa.  (Note that $\del_z^\dagger = \del_z$
and $\del_\zb^\dagger = \del_\zb$ because the hermitian conjugation is
defined with respect to Minkowski spacetime.)  The central charges of
$\Phi$ and $\Phib$ are given by their masses:
\begin{equation}
  Z(\Phi) = m \,,
  \qquad
  Z(\Phib) = -m \,.
\end{equation}

We consider a theory described by $n$ chiral multiplets
$\Phi^a = (\phi^a, \psi_\pm^a, F^a)$, $a = 1$, $\dotsc$, $n$, and their
conjugates $\Phib^\ab = (\phib^\ab, \psib_\pm^\ab, \Fb^\ab)$, with no
superpotential.  The action of the theory is
\begin{multline}
  \label{eq:S-3d-chiral}
  \int_{\R \times \C} \rmd^3 x
  (\Qb_- Q_- Q_+ \Qb_+) \cdot (g_{a\bb} \phi^a \phib^\bb)
  \\
  =
  \int_{\R \times \C}  \rmd^3 x \, g_{a\bb} \Bigl(
  \del_\zb\phi^a \del_z\phib^\bb
  + (\del_t + m_a)\phi^a  (\del_t + m_b)\phib^\bb
  - F^a \Fb^\bb
  \\  
  + \iu\bigl(\del_\zb\psi_+^a + (\del_t + m_a)\psi_-^a\bigr) \psib_+^\bb
  + \iu\psi_-^a \del_z\psib_-^\bb
  - i \psi_+^a (\del_t + m_b)\psib_-^\bb\Bigr) \,,
\end{multline}
where $g_{a\bb} = \delta_{ab}$.  We assign $(R,J) = (0,0)$ to $\Phi^a$.
Then, $\phi^a$, $\psi_\pm^a$ and $F^a$ have $(R,J) = (0,0)$,
$(-1,\pm 1/2)$ and $(-2,0)$, respectively, and $(-1)^F = (-1)^R$.

To deal with the $Q$-cohomology of this theory, it is better to switch
to a notation suitable for twisted theories.  Under the twisted
rotations on $\C$, all fields transform like components of
differential forms, which we can make manifest by renaming them as
follows:
\begin{equation}
  \label{eq:renamed-chiral}
  \begin{gathered}
  \chi^a = \iu\psi_+^a \rmd t - \iu\psi_-^a \rmd\zb \,,
  \qquad
  G^a = F^a \rmd t \wedge \rmd \zb \,,
  \\
  \etab^\ab = \psib_-^\ab \,,
  \qquad
  \xib_a = g_{a\bb} \psib_+^\bb \rmd z \,.
  \qquad
  \Gb_a = g_{a\bb} \Fb^\bb \rmd z \,.
  \end{gathered}
\end{equation}
The supercharge $Q = \Qb_-$ acts on the fields as
\begin{equation}
  \begin{alignedat}{2}
    Q \cdot \phi^a &= 0 \,,
    &
    Q \cdot \phib^\ab &= \etab^\ab \,,
    \\
    Q \cdot \chi^a &= \rmd'_{m_a} \phi^a \,,
    & \qquad
    Q \cdot \etab^\ab &= 0 \,,
    \\
    Q \cdot G^a &= \rmd'_{m_a} \chi^a \,,
    &
    Q \cdot \xib_a &= \Gb_a \,,
    \\
    &&
    Q \cdot \Gb_a &= 0 \,.
  \end{alignedat}
\end{equation}
Here
\begin{equation}
  \rmd'_m = \rmd t(\del_t + m) + \rmd\zb \del_\zb
\end{equation}
is the $\rmd'$-operator twisted by mass $m$.  The
action~\eqref{eq:S-3d-chiral} can be written as
\begin{equation}
  \int_{\R \times \C} Q \cdot
  \Bigl(
  g_{a\bb} \chi^a \wedge
  \star\bigl(\rmd t(\del_t + m_b) + \rmd z \del_z\bigr) \phib^\bb
  + 2\iu G^a \wedge \xib_a \Bigr) \,.
\end{equation}

The Poisson vertex algebra $\CV$ is the $Q$-cohomology taken in the
space of twisted-translated local operators.  In terms of these
operators the twisted masses disappear from the $Q$-action:
\begin{equation}
  Q \cdot {}^Z\chi^a = \rmd' {}^Z\phi^a \,,
  \qquad
  Q \cdot {}^ZG^a = \rmd' {}^Z\chi^a \,.
\end{equation}
It follows that the $Q$-cohomology is independent of the twisted
masses.  (The dependence of the action on the twisted masses is
$Q$-exact.)  This is a consequence of the locality of the Poisson
vertex algebra: the twisted masses may be thought of as vacuum
expectation values of scalars in nondynamical vector multiplets for
flavor symmetries and, as such, are determined by the boundary
condition at infinity, on which $\CV$ does not depend.  We will set
$m_a = 0$ below to simplify our argument.

Not all local operators are relevant for the computation of $\CV$.
Suppose that we rewrite the action in such a way that it couples to
the product metric on $\R \times \C$ in a manner that is invariant
under diffeomorphisms on $\R$ and those on $\C$.  (This is achieved by
introduction of the topological term
$\int_{\R \times \C} \rmd t \wedge \phi^* \omega_{\C^n}$ to the
action, where
$\omega_{\C^n} = \iu g_{a\bb} \rmd\phi^a \wedge \rmd\phib^\bb$ is the
K\"ahler form on $\C^n$.)  If $V$ is a vector field generating such
diffeomorphisms, $J_V = V^\mu T_{\mu\nu} \rmd x^\nu$ is a current such
that an integral of $\star J_V$ acts on local operators by $V$.  The
components $T_{tt}$ and $T_{\zb\zb}$ of the stress--energy tensor are
$Q$-exact because the action is $Q$-exact and variations of the
corresponding components of the metric commute with the $Q$-action,
while $T_{tz} = T_{t\zb} = 0$ as the metric is of a product form and
$T_{z\zb} = 0$ by conformal invariance on $\C$.  As a result, the
diffeomorphisms on $\R$ and antiholomorphic reparametrizations on $\C$
act on the $Q$-cohomology trivially.  In particular, $Q$-closed
operators that have nonzero scaling dimension on $\R$ or nonzero
antiholomorphic scaling dimension on $\C$ are $Q$-exact.

The relevant operators are therefore local operators constructed from
$\phi^a$, $\phib^\ab$, $\etab^\ab$, $\xib_a$ and $\Gb_a$, as well as
their $z$-derivatives.  Among these, $\Gb_a$ are auxiliary fields and
vanish by equations of motion.  Since
$g_{a\bb} \del_z \etab^\bb = \del_t \xib_{az}$ by an equation of
motion, $\del_z^k\etab^\ab$ with $k \geq 1$ can be dropped.  In the
absence of $\del_z^k \etab^\ab$ and $\del_t^k \xib_{az}$, operators
that contain $\del_z^k\phib^\ab$ cannot be $Q$-closed.  Thus we can
also drop $\del_z^k\phib^\ab$.

We may think of $(\phi^a, \phib^\ab)$ as coordinates on the target space
\begin{equation}
  X = \C^n
\end{equation}
and identify $\etab^\ab$ with $\rmd\phib^\ab$.  Furthermore, for each
$k \geq 1$, we may regard $\del_z^k\phi^a$ as a section of the
holomorphic cotangent bundle $T_X^\vee$, and $\del_z^{k-1} \xib_{az}$
as a section of the tangent bundle $T_X$.  Under this identification,
the relevant operators with $J' = j'$ are sections of
$\Omega^{0,\bullet}_X \otimes E^{j'}_X$, where $E^{j'}_X$ is a
holomorphic vector bundle given by the formal series
\begin{equation}
  \label{eq:E_X}
  \bigoplus_{J' = 0}^\infty u^{J'} E^{J'}_X
  =
  \bigotimes_{k = 1}^\infty
  \Biggl(
  \biggl(
  \bigoplus_{l = 0}^\infty u^{lk} S^l T^\vee_X
  \biggr)
  \otimes
  \biggl(
  \bigoplus_{m = 0}^\infty u^{mk}
  \Lambda^m T_X
  \biggr)
  \Biggr)
  \,.
\end{equation}
For each $k$, the factor of the $l$th symmetric tensor power
$S^l T^\vee_X$ accounts for operators that take the form
$f(\phi,\phib)_{a_1 \dotso a_l} \del_z^k \phi^{a_1} \dotsb \del_z^k
\phi^{a_l}$, and the $m$th exterior power $\Lambda^m T_X$ accounts for
those of the form
$f(\phi,\phib)^{a_1 \dotso a_m} \del_z^{k-1} \xib_{a_1 z} \dotsb
\del_z^{k-1} \xib_{a_m z}$.

Using the equations of motion $\Gb_a = 0$, we see that $Q$ acts on
these sections as the Dolbeault operator $\delb$, increasing $R$ by
$1$ but keeping $J'$ unchanged.  We conclude that the $Q$-cohomology
of local operator is isomorphic to the Dolbeault cohomology of this
bundle:
\begin{equation}
  \CV = \bigoplus_{J' = 0}^\infty \CV^{J'} \,,
  \qquad
  \CV^{J'} \iso H_\delb^\bullet(X; E^{J'}_X) \,.
\end{equation}
For $X = \C^n$, we have $H_\delb^q(X; E^{J'}_X) = 0$ for $q \geq 1$ by
the $\delb$-Poincar\'e lemma.  Therefore, $\CV$ is isomorphic as an
algebra to the algebra of holomorphic sections of $E^{\bullet}_X$.

The $\lambda$-bracket has $(R, J') = (-1,-1)$.  There are three
combinations of local operators for which we need to compute the
$\lambda$-bracket, namely
$\lbracket{\lambda}{\ec{\phi^a}}{\ec{\phi^b}}$,
$\lbracket{\lambda}{\ec{\xib_{az}}}{\ec{\xib_{bz}}}$ and
$\lbracket{\lambda}{\ec{\xib_{az}}}{\ec{\phi^b}}$; the other
combinations can be obtained from these by the Leibniz rule,
sesquilinearity and symmetry of the $\lambda$-bracket.

Since there are no $Q$-cohomology classes with $R < 0$, we immediately
find
\begin{equation}
  \lbracket{\lambda}{\ec{\phi^a}}{\ec{\phi^b}} = 0 \,.
\end{equation}
In general, $\lbracket{\lambda}{\ec{\CO_1}}{\ec{\CO_2}}$ vanishes
unless the $d$th descendant of $\CO_1$ produces a singularity when
placed at the same point as $\CO_2$, for otherwise the integration
cycle $S^{d+1}_{x_2}$ can be shrunk to a point.  The theory under
consideration is a free theory, so the product of a bosonic field and
a fermionic one, such as those that appear in
$\lbracket{\lambda}{\ec{\phi^a}}{\ec{\phi^b}}$, cannot be singular.
For the same reason we have
\begin{equation}
  \lbracket{\lambda}{\ec{\xib_{az}}}{\ec{\xib_{bz}}} = 0 \,.
\end{equation}

The remaining combination
$\lbracket{\lambda}{\ec{\xib_{az}}}{\ec{\phi^b}}$ is a polynomial in
$\lambda$ with $(R, J') = (0,0)$.  Since $J'(\lambda) = 1$ and there
are no local $Q$-cohomology classes with $J' < 0$, it must be actually
independent of $\lambda$ and equal to a $Q$-cohomology class with
$(R, J') = (0,0)$, that is, a holomorphic function of $\phi^a$.  The
theory has a $\U(n)$ global symmetry under which $\phi^a$ transform in
the vector representation and $\xib_a$ in the dual representation, and
the $\lambda$-bracket must be invariant under this symmetry.  These
constraints leave the only possibility to be that
$\lbracket{\lambda}{\ec{\xib_{az}}}{\ec{\phi^b}}$ is proportional to
$\delta_a^b$.

Let us calculate this $\lambda$-bracket explicitly.  For this
calculation we restore the twisted masses to demonstrate that they do
not affect the result.

The first descendant of $\xib_{az}$ is given by
\begin{equation}
  \Qone \cdot \xib_{az}
  = g_{a\bb} \bigl(\rmd t \del_z - \rmd\zb (\del_t + m_a)\bigr) \phib^\bb
  \,,
\end{equation}
so we have
\begin{equation}
  \lbracket{\lambda}{\ec{\xib_{az}}}{\ec{\phi^b}}(0)
  =
  -\biggec{\biggl(\int_{S^2_0}
  e^{\lambda z} \rmd z \wedge
  e^{-m_a t}
  g_{a\cb} \bigl(\rmd t \del_z - \rmd\zb (\del_t + m_a)\bigr) \phib^\cb
  (x) \biggr)
  \phi^b(0)}
\end{equation}
where the factor of $e^{-m_a t}$ comes from twisted translation.  Using
the Stokes theorem we can convert this integral to one over the
$3$-ball $B^3_0$ that bounds $S^2_0$:
\begin{equation}
  \label{eq:xiphi-int}
  -\frac{\iu}{2} \biggec{\biggl(\int_{B^3_0} \rmd^3 x
  \,
  e^{\lambda z - m_a t}
  g_{a\cb} (\del_t^2 + \del_z \del_\zb - m_a^2) \phib^\cb(x) \biggr)
  \phi^b(0)} \,.
\end{equation}
By integration by part in the path integral we obtain the identity
\begin{equation}
  \frac{\delta\phi^b(0)}{\delta\phi^a(x)}
  =
  \frac{\delta S}{\delta\phi^a(x)} \phi^b(0) \,,
\end{equation}
which holds inside correlation functions as long as no other operators
are present at $x$.  The left-hand side is $\delta^b_a$ times the
delta function supported at $x = 0$, whereas the right-hand side is
$-g_{a\cb} (\del_t^2 + \del_z \del_\zb - m_a^2) \phib^\cb(x)
\phi^b(0)$.  Plugging this relation into the
integral~\eqref{eq:xiphi-int}, we find
\begin{equation}
  \lbracket{\lambda}{\ec{\xib_{az}}}{\ec{\phi^b}}(0)
  =
  \frac{\iu}{2} \delta_a^b
  \,.
\end{equation}
As expected, the result is proportional to $\delta_a^b$, with the
proportionality constant independent of the twisted masses.

Finally, let us consider the Poisson algebra $\CV/(\del_z\CV) \CV$
associated with the dimensional reduction of the theory.  The elements
of $\CV/(\del_z\CV) \CV$ are in one-to-one correspondence with the
local operators constructed from $\phi^a$ and $\xib_{az}$ but not
their $z$-derivatives, that is, the holomorphic sections of
$\Lambda^\bullet T_X$.  The Poisson bracket on $\CV/(\del_z\CV) \CV$
coincides with the Schouten--Nijenhuis bracket.  This Poisson algebra
is the one for the B-model with target space $X$~\cite{Beem:2018fng}.

\subsection{Sigma models}

The construction of the Poisson vertex algebra discussed above can be
generalized to the case when the target space $X$ is not $\C^n$ but
any other K\"ahler manifold.  Thus we get a map
\begin{equation}
  X \mapsto \CV(X)
\end{equation}
which assigns to a K\"ahler manifold $X$ a Poisson vertex algebra
$\CV(X)$.  Classically, $\CV(X)$ is still described by the Dolbeault
cohomology of $X$ with values in the holomorphic vector
bundles~\eqref{eq:E_X}.  Quantum corrections may alter this
description, however.

\subsection{Free vector multiplets}

A theory of free vector multiples provides an example of a sigma model
with $X \neq \C^n$.  In three dimensions, an abelian gauge field $A$
can be dualized to a periodic scalar $\gamma$ through the relation
$\rmd\gamma = \iu \star \rmd A$ (in Euclidean signature).  Under this
dualization process a vector multiplet is mapped to a chiral multiplet
whose scalar field is $\phi = \sigma + \iu\gamma$, where $\sigma$ is
the adjoint scalar in the vector multiplet.  Hence, a theory of $n$
abelian vector multiplets is equivalent to a sigma model with target
$X = (\R \times S^1)^n$.

The target space metric enters the action through $Q$-exact terms, so
we can actually take
\begin{equation}
  X = (\C^\times)^n \,,
\end{equation}
with the standard flat metric on it.  The target being flat, the
theory is free and there are no quantum corrections to the description
of the Poisson vertex algebra as the Dolbeault cohomology.

Compared to the case of $X = \C^n$, we have more ingredients to build
$Q$-cohomology classes from: holomorphic sections of the bundles
$E_X^\bullet$ can have poles at $\phi^a = 0$, and $\etab^\ab/\phib^\ab$
are $Q$-closed but not $Q$-exact.  Accordingly, we have more
combinations to consider for the $\lambda$-bracket.

Specifically, we have the cases when either operator in the
$\lambda$-bracket involves $\ec{\etab^\ab/\phib^\ab}$, and need to
evaluate $\lbracket{\lambda}{\ec{\phi^a}}{\ec{\etab^\bb/\phib^\bb}}$,
$\lbracket{\lambda}{\ec{\etab^\ab/\phib^\ab}}{\ec{\etab^\bb/\phib^\bb}}$
and $\lbracket{\lambda}{\ec{\xib_{az}}}{\ec{\etab^\bb/\phib^\bb}}$.
However, these additional combinations all vanish.  For the first two,
this is because there are simply no $Q$-cohomology classes with
$J' = -1$.  The last one has $(R, J') = (1,0)$ and may be proportional
to $\ec{\etab^\ab/\phib^\ab}$.  This vanishes since neither
$\etab^\ab$ nor $\phib^\ab$ produces a singularity when multiplied by
$\xib_{az}^{(1)}$.

\subsection{\texorpdfstring{$\CN = 2$ superconformal field
    theories}{N = 2 superconformal field theories}}

Although the Poisson vertex algebra of an interacting $\CN = 2$
supersymmetric field theory is difficult to determine, some general
statements can be made about it if the theory is unitary and has
conformal symmetry.

In that case, by conformal symmetry the $Q$-cohomology of local
operators is isomorphic as a vector space to the $Q$-cohomology of
states in radial quantization, and by unitarity the latter is
isomorphic to the space of $Q$-harmonic states on which
$\{Q, Q^\dagger\} = 0$.  In radial quantization, the hermitian
conjugate of $Q$ is a conformal supercharge and satisfies
\begin{equation}
  [Q, Q^\dagger] = D - R - J \,,
\end{equation}
where $D$ is the dilatation operator.  Hence, a local $Q$-cohomology
class is represented uniquely by a local operator with
$D - R - J = 0$.  If we assign $D(\lambda) = 1$, the $\lambda$-bracket
has $(D,R,J) = (-3/2,-1,-1/2)$ and preserves $D - R - J$.

There are many $\CN = 2$ superconformal multiplets that contain such
local operators; see~\cite{Cordova:2016emh} for a comprehensive list.
Especially important are those with conserved currents.

For example, there is a flavor current multiplet which contains a
fermionic operator $\Psi_0$ with $(D,R,J) = (3/2,1,1/2)$.  The first
descendant $\Psi_0^{(1)}$ of this operator is given by
$\rmd z \wedge \Psi_0^{(1)} = \star (j - j_\zb \rmd\zb)$, where $j$ is
the conserved current for a flavor symmetry.  The component $j_\zb$ is
$Q$-exact, so the zero mode $\zmode\ec{\Psi_0}$ acts on $\CV$ by an
infinitesimal flavor transformation.  For the theory of free chiral
multiplets, $(\Psi_0)^a{}_b = \xib_{bz}\phi^a$.

Similarly, for each integer $n \geq 1$, there is a multiplet that
contains conserved currents and a local operator $\Psi_n$ with
$(D,R,J) = (n/2 + 3/2,1,n/2 + 1/2)$ representing a $Q$-cohomology
class.  For $n = 1$, the first descendant $\Psi_1^{(1)}$ is part of a
spin-$3/2$ current, and $\zmode\ec{\Psi_1}$ acts by a supersymmetry
transformation.  The multiplet with $n = 2$ contains a stress--energy
tensor, and $\zmode\ec{\Psi_2}$ acts by the holomorphic derivative
$\del_z$.  For free chiral multiplets,
$(\Psi_2)^a{}_b = \xib_{bz} \del_z\phi^a$.  The multiplets with
$n \geq 3$ contain higher spin currents.

\section{\texorpdfstring{$\CN = 2$ supersymmetric field theories in four
    dimensions}{N = 2 supersymmetric field theories in four
    dimensions}}
\label{sec:4d}

Next, we turn to Poisson vertex algebras for four-dimensional
supersymmetric field theories.  In four dimensions, we must have twice
as many topological directions as in three dimensions.  It turns out
that we also need twice as many supercharges, namely eight
supercharges, generating $\CN = 2$ supersymmetry.  We will take
\begin{equation}
  M \times C = \R^2 \times \C \,.
\end{equation}

\subsection{Topological--holomorphic sector}
\label{sec:PVA-4d}

The supercharges $Q^A_\alpha$, $\Qb_{A\dot\alpha}$, $A = 1$, $2$,
$\alpha = \pm$, of the $\CN = 2$ supersymmetry algebra in $3+1$
dimensions satisfy the commutation relations
\begin{align}
  [Q^A_\alpha, \Qb_{B\dot\alpha}]
  &= \sigma_{\alpha\dot\alpha}^\mu P_\mu \delta^A_B \,,
  \\
  [Q^A_\alpha, Q^B_{\beta}]
  &= \epsilon_{\alpha\beta} \epsilon^{AB} Z \,,
  \\
  [\Qb_{A\dot\alpha}, \Qb_{B\dot\beta}]
  &= -\epsilon_{\dot\alpha\dot\beta} \epsilon_{AB} \Zb \,,
\end{align}
where $\Zb = Z^\dagger$ is a complex central charge.  Our convention
is such that
\begin{equation}
  \sigma^0
  =
  \biggl(
  \begin{array}{cc}
    -1 & 0 \\
    0 & -1
  \end{array}
  \biggr)
  \,,
  \qquad
  \sigma^1
  =
  \biggl(
  \begin{array}{cc}
    0 & 1 \\
    1 & 0
  \end{array}
  \biggr)
  \,,
  \qquad
  \sigma^2
  =
  \biggl(
  \begin{array}{cc}
    0 & -\iu \\
    \iu & 0
  \end{array}
  \biggr)
  \,,
  \qquad
  \sigma^3
  =
  \biggl(
  \begin{array}{cc}
    1 & 0 \\
    0 & -1
  \end{array}
  \biggr)
\end{equation}
and
$\epsilon^{12} = -\epsilon_{12} = -\epsilon_{+-} =
-\epsilon_{\dot+\dot-} = 1$.  We raise and lower indices as
$\Qb^A_{\dot\alpha} = \epsilon^{AB} \Qb_{B\dot\alpha}$,
$\Qb_{A\dot\alpha} = \epsilon_{AB} \Qb^B_{\dot\alpha}$.

We perform the Wick rotation $x^0 \to -\iu x^4$ and introduce the
complex coordinates $w = (x^1 + \iu x^2)/2$ and
$z = (x^3 + \iu x^4)/2$.  The generators
$P_{\alpha\dot\alpha} = \sigma_{\alpha\dot\alpha}^\mu P_\mu$ of
translations are then given by
\begin{equation}
  \biggl(
  \begin{array}{cc}
    P_{+\dot+} & P_{+\dot-} \\
    P_{-\dot+} & P_{-\dot-}
  \end{array}
  \biggr)
  =  
  \biggl(
  \begin{array}{cc}
    P_z & P_w \\
    P_\wb & -P_\zb
  \end{array}
  \biggr) \,.
\end{equation}
We choose the $w$-plane for $M = \R^2$ and the $z$-plane for $C = \C$.

We need a supercharge $Q$ such that $Q^2 = 0$ and $P_{+\dot- }$,
$P_{-\dot+}$, $P_{-\dot-}$ are $Q$-exact.  Since $P_z$ should not
enter any $Q$-commutators, $Q$ must be of the form
\begin{equation}
  Q = a Q^1_- + b Q^2_- - c \Qb_{2\dot-} + d \Qb_{1\dot-} \,.
\end{equation}
We have
\begin{equation}
  Q^2 = -\det(A) P_\zb \,,
  \qquad
  A
  =
  \biggl(
  \begin{array}{cc}
    a & b \\
    c & d
  \end{array}
  \biggr)
  \,.
\end{equation}
For $Q^2 = 0$, the two rows of the matrix $A$ must be proportional to
each other.  On the other hand, the $Q$-commutators of supercharges
that contain $P_{+\dot-}$ and $P_{-\dot+}$ can be written as
\begin{equation}
\label{eq:Q-P1-P2}
  [Q,
  \biggl(
  \begin{array}{cc}
    Q^2_+ & -Q^1_+ \\
    \Qb_{1\dot+} & \Qb_{2\dot+}
  \end{array}
  \biggr)
  ]
  =
  \biggl(
  \begin{array}{cc}
    Z & -P_w \\
    P_\wb & \Zb
  \end{array}
  \biggr)
  A
  \,.
\end{equation}
(The commutator on the left-hand side is to be understood as acting on
each entry of the matrix.)  For $P_w$ and $P_\wb$ to both appear on
the right-hand side, neither row of $A$ should be zero.  Thus we find
\begin{equation}
  \label{eq:Qab}
  Q = a(Q^1_- + t\Qb_{2\dot-}) + b(Q^2_-  - t\Qb_{1\dot-})
\end{equation}
for some $t \neq 0$.

From the commutators~\eqref{eq:Q-P1-P2} we see that the linear
combinations
\begin{equation}
  \label{eq:P_w^Z}
  P_w + t^{-1} Z \,,
  \qquad
  P_\wb - t \Zb
\end{equation}
are $Q$-exact.  As in the case of three-dimensional $\CN = 2$
supersymmetry, we define twisted-translated local operators by
\begin{equation}
  \begin{split}
    {}^{t,Z}\CO(w,\wb,z,\zb)
    &= \exp\bigl(\iu(P_w + t^{-1} Z)w + \iu(P_\wb - t\Zb)\wb\bigr)
       \cdot \CO(0,0, z,\zb)
    \\
    &= \exp(\iu t^{-1} Zw - \iu t\Zb\wb) \cdot \CO(w,\wb, z,\zb)
  \end{split}
\end{equation}
so that $\iu(P_w + t^{-1} Z)$ and $\iu(P_\wb - t\Zb)$ act on
${}^{t,Z}\CO$ as $\del_w$ and $\del_\wb$, respectively.
Twisted-translated local $Q$-cohomology classes form a Poisson vertex
algebra $\CV$.%
\footnote{The fact that the $Q$-cohomology of local operators is a
  Poisson vertex algebra was mentioned in
  \cite{Beem:String-Math-2017,Beem:Pollica,Beem:String-Math-2019}.}

Let us suppose that the theory has an R-symmetry $\SU(2)_R$ under
which $(Q^1_\alpha, Q^2_\alpha)$ and
$(\Qb^1_{\dot\alpha}, \Qb^2_{\dot\alpha})$ transform as doublets.
Then, the structure of $\CV$ does not depend on the parameters $a$,
$b$ even though they appear in the formula~\eqref{eq:Qab} for $Q$.
Under the action of the complexification $\SL(2,\C)_R$ of the
R-symmetry group, $\CV$ gets mapped to an isomorphic algebra.  Since
the linear combinations of supercharges in the parentheses in the
expression~\eqref{eq:Qab} form a doublet of $\SU(2)_R$, by the
transformation by
\begin{equation}
  \biggl(
  \begin{array}{cc}
    0 & -b \\
    b^{-1} & a
  \end{array}
  \biggr)
  \in
  \SL(2,\C)_R  
\end{equation}
we can set
\begin{equation}
  Q = Q^1_- + t\Qb_{2\dot-}
\end{equation}
(assuming $b \neq 0$ without loss of generality).  With this choice,
$Q$ has charge $R = 1$ under the diagonal subgroup $\U(1)_R$ of
$\SU(2)_R$.
%
% Our $R$ is twice $\CR$ of Beem et al.
%
Under the rotation symmetry on $\C$, it has spin $J(Q) = -1/2$.

Also, $\CV$ is independent of $t$ if the theory has rotation symmetry
on $\R^2$.  Since $Q^1_-$ and $\Qb_{2\dot-}$ have spin $\Jt = -1/2$
and $1/2$ under the rotation symmetry, we can change the value of $t$
using complexified rotations; in effect, $t$ transforms as if it has
weight $1$ under this $\C^\times$-action, just as $\del_w$ does.
Since rotations act on the space of local operators at $w = 0$, it
follows that the $Q$-cohomology of local operators, as a vector space,
is independent of $t$ at $w = 0$, hence anywhere on $\R^2$.  Moreover,
by the action of the rotation symmetry, we can also show that the
algebra structure and the $\lambda$-bracket remain the same under
phase rotation of $t$.%
\footnote{Here we cannot use the $\C^\times$-action because we need to
  place local operators away from $w = 0$ in order to define these
  structures.  The action by $\alpha \in \C^\times$ transforms
  ${}^{t,Z}\CO(w,\wb,z,\zb)$ to
  ${}^{\alpha t,Z}\CO(\alpha w, \alpha^{-1}\wb,z,\zb)$ (assuming, for
  simplicity, that $\CO$ is a scalar operator).  For $\alpha w$ and
  $\alpha^{-1} \wb$ to be complex conjugate to each other, we must
  have $\alpha \in \U(1)$ or $w = 0$.}
These structures depend holomorphically on $t$, so they are actually
entirely independent of $t$.  Thus, we can take
\begin{equation}
  t = 1
\end{equation}
and
\begin{equation}
  \label{eq:Q}
  Q = Q^1_- + \Qb_{2\dot-} \,.
\end{equation}
We will write ${}^Z\CO$ for ${}^{1,Z}\CO$:
\begin{equation}
  \begin{split}
    {}^Z\CO(w,\wb,z,\zb)
    = \exp(\iu Zw - \iu\Zb\wb) \cdot \CO(w,\wb, z,\zb)
  \end{split}
\end{equation}

The one-form supercharge is not uniquely determined.  The general form
of $\Qone$ that has $(R,J) = (-1, 1/2)$ is
\begin{equation}
  \label{eq:Qone}
  \Qone
  =
  \iu Q^2_+ \rmd w + \iu \Qb_{1\dot+} \rmd\wb
  + \iu \bigl((u - 1) Q^2_- - u\Qb_{1\dot-}\bigr) \rmd\zb \,,
\end{equation}
with $u$ being an arbitrary complex number.  The $\lambda$-bracket
does not depend on $u$, however.  To see this, pick a boundary
condition such that all local $Q$-cohomology classes vanish at the
infinity of $\R^2$.  Furthermore, we choose to represent the homology
class $\ec{S^3_{x_2}}$ in the definition~\eqref{lbracket} of the
$\lambda$-bracket by a ``cylinder'' whose ``side'' is
$\R^2 \times S^1_{z_2} \subset \R^2 \times \C$, where $S^1_{z_2}$ has
a fixed radius everywhere except at the infinity of $\R^2$ where it
shrinks to a point; by the Poincar\'e conjecture, this cylinder is
homeomorphic to $S^3_{x_2}$.  Then, $\Qone_\zb$ drops out of the
computation of the $\lambda$-bracket because the only contributions
come from the region in which the pullback of $\rmd z \wedge \rmd\zb$
vanishes.

\subsection{\texorpdfstring{Vertex algebras for $\CN = 2$
    superconformal field theories}{Vertex algebras for N = 2
    superconformal field theories}}

If the theory has not only $\CN = 2$ supersymmetry but also conformal
symmetry, $\CV$ can be deformed to a family of vertex algebras
$V^\hbar$ parametrized by $\hbar \in \C$.  This is essentially
quantization of $\CV$ by
$\Omega$-deformation~\cite{Oh:2019bgz,Jeong:2019pzg}, and related via
dimensional reduction to the quantization of a Poisson algebra
associated with an $\CN = 4$ superconformal field theory in three
dimensions~\cite{Beem:2016cbd,Beem:2018fng}.

An $\CN = 2$ superconformal field theory has eight conformal
supercharges $S_A^\alpha$, $\Sb^A_{\dot\alpha}$, in addition to the
Poincar\'e supercharges $Q^A_\alpha$, $\Qb_{A\dot\alpha}$.  In radial
quantization the two sets of supercharges are related by hermitian
conjugation:
\begin{equation}
  (Q^A_\alpha)^\dagger = S_A^\alpha \,,
  \qquad
  (\Qb_{A\dot\alpha})^\dagger = \Sb^{A\dot\alpha} \,.
\end{equation}
Furthermore, the theory has an extra R-symmetry $\U(1)_r$, under which
$Q^A_\alpha$, $\Sb^A_{\dot\alpha}$ have charge $r = 1/2$ and
$\Qb_{A\dot\alpha}$, $S_A^\alpha$ have $r = -1/2$.  The central charge
$Z$ necessarily vanishes because of $\U(1)_r$.

Let us introduce the following deformations of $P$, $Q$ and $\Qone$:
\begin{align}
  P^\hbar &= P_w \rmd w + P_\wb \rmd\wb + P_z \rmd z
             + (P_\zb - \hbar R^2{}_1) \rmd\zb \,, \\
  Q^\hbar &= Q^1_- + \Qb_{2\dot-} + \hbar(\Sb^{2\dot-} - S_1^-) \,, \\
  \Qone^\hbar &= \Qone \,,
\end{align}
Here $R^A{}_B$ are the generators of
$\SU(2)_R \times \U(1)_r \iso \U(2)$, in terms of which
$R = R^1{}_1 - R^2{}_2$ and $r = R^1{}_1 + R^2{}_2$.  These deformed
generators satisfy the relations
\begin{align}
  \label{eq:Q2-Omega}
  (Q^\hbar)^2 &= \hbar (\Jt + r) \,,
  \\
  \label{eq:PQ-Omega}
  [Q^\hbar, P^\hbar_\mu] &= -\iu\hbar \del_\mu \Jt^\nu \Qone^\hbar_\nu \,,
  \\
  \label{eq:PQone-Omega}
  [\Qone^\hbar, P^\hbar_\mu] &= 0 \,,
  \\
  [Q^\hbar, \Qone^\hbar]
  &= \iu P^\hbar_i \, \rmd y^i + \iu P^\hbar_\zb \, \rmd\zb \,.
\end{align}
Recall that $\Jt$ is the generator of rotations on $\R^2$;
% 
% $\Jt = \CM_+{}^+ - \CM^{\dot+}{}_{\dot+}$ of Beem et al.
%
it is normalized in such a way that $[\Jt, P_w] = P_w$ and
$[\Jt, P_\wb] = -P_\wb$.  We have also denoted the corresponding
vector field by the same symbol:
\begin{equation}
  \Jt = w \del_w - \wb \del_\wb \,.
\end{equation}
%
% Note the sign.  We can confirm this formula by letting $[\Jt, P_w]$
% act on fields.
%
We regard $P^\hbar$ as a twisted translation generator and define
twisted-translated local operators by
\begin{equation}
  \begin{split}
    \CO^\hbar(w,\wb,z,\zb)
    = \exp(\iu z P^\hbar_z + \iu\zb P^\hbar_\zb) \cdot \CO(w,\wb, 0,0)
  \end{split}
\end{equation}
so that $\iu P^\hbar_\mu$ act on $\CO^\hbar$ as $\del_\mu$.

The deformed supercharge $Q^\hbar$ does not square to zero, but
instead to the generator
\begin{equation}
  \Jt' = \Jt + r
\end{equation}
of twisted rotations on $\R^2$.  (In this sense $Q^\hbar$ defines an
$\Omega$-deformation~\cite{Nekrasov:2002qd,Nekrasov:2003rj} and
induces quantization~\cite{Yagi:2014toa}.)  We can still consider the
$Q^\hbar$-cohomology in the space of operators that are annihilated by
$\Jt'$.  The vertex algebra $V^\hbar$ is the $Q^\hbar$-cohomology of
twisted-translated local operators, placed at $w = 0$ and annihilated
by $\Jt'$.  Since $P^\hbar_\zb$ is $Q^\hbar$-exact, classes in
$V^\hbar$ vary holomorphically on $\C$, just as in the undeformed
case.  Unlike the undeformed case, however, the product
$\ec{\CO_1^\hbar(z_1)} \ec{\CO_2^\hbar(z_2)}$ of two local
$Q$-cohomology classes can be singular at $z_1 = z_2$ because
operators representing these classes are located at the same point
$w = 0$ on $\R^2$.

The structure of $V^\hbar$ is most naturally described in the Kapustin
twist~\cite{Kapustin:2006hi}, in which the rotation generators
$(\Jt, J)$ are replaced by $(\Jt', J')$, where $J' = J + R/2$ as in
the case of three-dimensional $\CN = 2$ supersymmetry.  In the
Kapustin twist, $Q$ transforms under rotations as a scalar and
$P^\hbar$, $\Qone$ as one-forms.  The $R$-grading is well defined in
$V^\hbar$ if we assign
\begin{equation}
  R(\hbar) = 2 \,,
\end{equation}
for then $Q^\hbar$ has a definite charge $R(Q^\hbar) = 1$, while
$R(P^\hbar) = 0$ and the twisted translation preserves $R$.  The
operator product expansion (OPE) takes the form
\begin{equation}
  \label{eq:V-OPE}
  \ec{\CO_a^\hbar(z_1)} \ec{\CO_b^\hbar(z_2)}
  =
  \sum_c
  \frac{C_{ab}{}^c(\hbar) \ec{\CO_c^\hbar(z_2)}}{(z_1 - z_2)^{h_a + h_b - h_c}}
  \,,
\end{equation}
where $h_a = J'(\CO_a)$ and $C_{ab}{}^c(\hbar)$ are analytic functions
of $\hbar$.  Having a singular OPE structure, $V^\hbar$ is a vertex
algebra, rather than a Poisson vertex algebra.

For a unitary $\CN = 2$ superconformal field theory, it was shown
in~\cite{Oh:2019bgz} that $V^\hbar$ is isomorphic to the vertex
operator algebra introduced in~\cite{Beem:2013sza}.  The latter is the
cohomology of twisted-translated local operators with respect to the
supercharge $Q^1_- + \hbar \Sb^{2\dot-}$, which is ``half'' of
$Q^\hbar$, so to speak.  This supercharge has $r = 1/2$, so $V^\hbar$
has an additional grading by $r$ in the unitary case.

We can encode the singular part of the OPE in the $\lambda$-bracket
$\lBracket{\lambda}{\ }{\ }\colon V^\hbar \otimes V^\hbar \to
V^\hbar[\lambda]$ on $V^\hbar$, defined by
\begin{equation}
  \begin{split}
    \lBracket{\lambda}{\ec{\CO_1^\hbar}}{\ec{\CO_2^\hbar}}(z_2)
    &=
    \int_{S^1_{z_2}} \frac{\rmd z_1}{2\pi\iu} e^{\lambda(z_1 - z_2)}
    \ec{\CO_1^\hbar(z_1)} \ec{\CO_2^\hbar(z_2)} \\
    &=
    \sum_{n=0}^\infty \frac{\lambda^n}{n!}
    \sum_c \delta_{h_1 + h_2 - h_c, n+1}
    C_{12}{}^c(\hbar) \ec{\CO_c^\hbar(z_2)} \,.
  \end{split}
\end{equation}
The $\lambda$-bracket may be considered for any vertex algebra, but
that of $V^\hbar$ has a special property.  In the limit $\hbar \to 0$,
nontrivial $Q^\hbar$-cohomology classes reduce to nontrivial
$Q$-cohomology classes; the $\hbar$-correction to $Q$ can destroy but
not create cohomology classes, as we will argue shortly.  Since the
OPE between $Q$-cohomology classes is regular, the OPE coefficients
$C_{ab}{}^c(\hbar)$ for $h_a + h_b - h_c > 0$ contain only terms of
positive order when expanded in powers of $\hbar$.  In particular, the
$\lambda$-bracket is at least of order $\hbar$.%
\footnote{In~\cite{Oh:2019bgz}, the construction of vertex algebras
  was extended to $\CN = 2$ supersymmetric gauge theories which are
  not necessarily conformal.  For a nonconformal theory, the
  associated vertex algebra is anomalous, but the anomaly can be
  canceled if an $\CN = (0,2)$ supersymmetric surface defect carrying
  an appropriate vertex algebra is inserted at $w = 0$.  Although the
  combined vertex algebra is well-defined, it may not have a classical
  limit since the vertex algebra on the surface defect may not have
  one.  What goes wrong in the above argument is the assertion that
  the OPE between $Q$-cohomology classes is regular.}

It is known that in such a situation, the family of vertex algebras
$V^\hbar$ reduces in the limit $\hbar \to 0$ to a Poisson vertex
algebra $\CV'$; see \cite{MR3751122} for a proof.  The
$\lambda$-bracket of $\CV'$ is given by
\begin{equation}
  \lbracket{\lambda}{\ec{\CO_1}}{\ec{\CO_2}}
  = \lim_{\hbar \to 0} \frac{1}{\hbar}
  \lBracket{\lambda}{\ec{\CO_1^\hbar}}{\ec{\CO_2^\hbar}} \,.
\end{equation}
The appearance of Poisson vertex algebras in this context was
previously noted in~\cite{Beem:2017ooy,Beem:2019tfp}.

Now we show that $\CV'$ is a subalgebra of the Poisson vertex algebra
$\CV$ associated with the topological--holomorphic sector of the
theory.

First of all, we note that it makes sense to compare $V^\hbar$ and
$\CV$ as vector spaces.  Since $\Jt'$ is $Q$-exact, for the
computation of the $Q$-cohomology of local operators we can restrict
$Q$ to the kernel of $\Jt'$.  We can also compute it anywhere on
$\R^2$.  Thus we can compute $\CV$, as a vector space, as the
$Q$-cohomology in the space of of local operators that have $\Jt' = 0$
and are located at $w = 0$.  This is the space in which the
$Q^\hbar$-cohomology of local operators is defined.

An injection $V^\hbar \to \CV$ is constructed as follows.  Let
$\ec{\CO^\hbar}$ be a nontrivial $Q^\hbar$-cohomology class.  We may
assume that $\ec{\CO^\hbar}$ neither vanishes nor diverges in the
limit $\hbar \to 0$; we can multiply $\ec{\CO^\hbar}$ by an
appropriate power of $\hbar$ if needed.  Thus we have
$\CO^\hbar = \CO + \hbar\CO_1$, with $\CO$ containing no $\hbar$ and
$\CO_1$ nonnegative powers of $\hbar$.  If we write
$Q^\hbar = Q + \hbar Q_1$, then the relation $(Q^\hbar)^2 = 0$ (which
holds in the space of operators under consideration) implies
$[Q,Q_1] = Q_1^2 = 0$, while $Q^\hbar \cdot \CO^\hbar = 0$ implies
$Q \cdot \CO = Q_1 \cdot \CO + Q \cdot \CO_1 + \hbar Q_1 \cdot \CO_1 =
0$.  Hence, $\CO$ represents a $Q$-cohomology class.  This class is
nontrivial because if $\CO = Q \cdot \CO'$ for some local operator
$\CO'$, then $Q^\hbar \cdot (\CO_1 - Q_1 \cdot \CO') = 0$ and
$\CO^\hbar = Q^\hbar \cdot \CO' + \hbar(\CO_1 - Q_1 \cdot \CO')$, in
contradiction with the assumption.

To show that the $\lambda$-bracket of $V^\hbar$ reduces to that of
$\CV$ in the limit $\hbar \to 0$, we introduce an equivariant analog
of topological--holomorphic descent.  The key equation~\eqref{eq:QO*}
is deformed to
\begin{multline}
  \label{eq:QO*-Omega}
  Q^\hbar \cdot \bigl(\CO^\hbar(w,\wb,z,\zb)\bigr)^{\pm*}
  = \pm(\rmd' + \hbar\iota_{\Jt})
    \bigl(\CO^\hbar(w,\wb,z,\zb)\bigr)^{\pm*}
  \\
  + \bigl(\exp(\iu w P^\hbar_w + \iu\wb P^\hbar_\wb) \cdot
    (Q^\hbar \cdot \CO(0,0,z,\zb))\bigr)^{\mp*} \,,
\end{multline}
For a local operator $\CO$ that is $Q^\hbar$-closed at $w = 0$, we
have
\begin{equation}
  Q^\hbar \cdot (\omega \wedge (\CO^\hbar)^*)
  = -\rmd^\hbar (\omega \wedge (\CO^\hbar)^*)
\end{equation}
for any holomorphic one-form $\omega$ on $C$, where
\begin{equation}
  \rmd^\hbar = \rmd + \hbar\iota_{\Jt} \,.
\end{equation}
Compared to the relation~\eqref{eq:QomegaO*}, the exterior derivative
$\rmd$ is deformed to its equivariant version $\rmd^\hbar$, which
squares to the Lie derivative by $\hbar\Jt$.

Accordingly, we can integrate $\omega \wedge (\CO^\hbar)^*$ over an
equivariant cycle $\Gamma^\hbar$ to obtain a $Q^\hbar$-closed
operator, and the $Q^\hbar$-cohomology class of the resulting operator
depends only on $\ec{\CO^\hbar} \in V^\hbar$ and the equivariant
homology class $\ec{\Gamma^\hbar}$.  The equivariant homology is the
homology with respect to the boundary operator
\begin{equation}
  \del^\hbar = \del + \hbar\CJ \,,
\end{equation}
where $\CJ$ is the dual of $\iota_\Jt$.  Acting on a $k$-chain located
at some fixed value of $\varphi = \arg w$, the operator $\CJ$ sweeps
it around in the $\varphi$-direction to produce a $(k+1)$-chain that
is invariant under the action of $\Jt = -\iu\del_\varphi$.

Let us consider the $Q^\hbar$-cohomology class
\begin{equation}
  \label{lbracket-hbar}
  \biggec{\biggl(\int_{S^3_{x_2}}
  e^{\lambda(z_1 - z_2)} \rmd z_1 \wedge (\CO_1^\hbar)^*(x_1) \biggr)
  \CO_2^\hbar(x_2)} \,.
\end{equation}
There is a $2$-disk $D^2_{x_2}$ such that
$\del^\hbar D^2_{x_2} = S^1_{z_2} + \hbar S^3_{x_2}/2\pi\iu$, with
$S^1_{z_2}$ being a circle lying in $\C$, centered at $z_2$ and located
at $w = 0$.%
\footnote{In terms of the spherical coordinates
  $(\psi,\theta,\varphi) \in [0,\pi] \times [0,\pi] \times [0,2\pi)$,
  defined by
  \begin{equation}
    \begin{aligned}
      x^1_1 - x^1_2 &= r \sin\psi \sin\theta \cos\varphi \,, \\
      x^2_1 - x^2_2 &= r \sin\psi \sin\theta \sin\varphi \,, \\
      x^3_1 - x^3_2 &= r \cos\psi \,, \\
      x^4_1 - x^4_2 &= r \sin\psi \cos\theta \,,
    \end{aligned}
  \end{equation}
  the $2$-disk $D^2_{x_2}$ is located at $\varphi = 0$ and has the
  boundary $S^1_{z_2}$ at $\theta = 0$, $\pi$.  One can easily show
  that for an equivariantly closed form, the integral over $S^3_{x_2}$
  reduces to an integral over $S^1_{z_2}$.}
Hence, we have
\begin{equation}
  \ec{S^3_{x_2}} = -\frac{2\pi\iu}{\hbar} \ec{S^1_{z_2}}
\end{equation}
in the equivariant homology, and the above $Q^\hbar$-cohomology class
is equal, up to an overall numerical factor, to
$\lBracket{\lambda}{\ec{\CO_1^\hbar}}{\ec{\CO_2^\hbar}}(z_2)/\hbar$.  In the
limit $\hbar \to 0$, the expression~\eqref{lbracket-hbar} reduces to
the $\lambda$-bracket $\lbracket{\lambda}{\ec{\CO_1}}{\ec{\CO_2}}(z_2)$,
leading to the desired relation between the two $\lambda$-brackets.

If the theory is unitary, the limit $\hbar \to 0$ of $V^\hbar$ equals
$\CV$ and not a proper subalgebra thereof.  In this case, the elements
of $V^\hbar$ are in one-to-one correspondence with the harmonic states
of $Q^\hbar$, that is, those states satisfying the condition
\begin{equation}
  [Q^\hbar, (Q^\hbar)^\dagger]
  = (1 + \hbar^2) (D - J - R)
  = 0 \,,
\end{equation}
where $D$ is the dilatation operator.  This condition is independent
of $\hbar$, hence equivalent to the harmonic condition
$[Q, Q^\dagger] = 0$ characterizing the elements of $\CV$.

The vacuum character of $V^\hbar$ computes the Schur limit of the
superconformal index of the theory \cite{Beem:2013sza}.  Since the
character is independent of $\hbar$, the character of $\CV$ also
equals the Schur index.  The Poisson vertex algebra can be defined for
nonconformal theories, and its character may be taken as a definition
of the Schur index for those theories.

\subsection{Free hypermultiplets}
\label{sec:free-hyper}

Let us determine the Poisson vertex algebra for the theory of a free
hypermultiplet.  To this purpose it is convenient to adopt a
two-dimensional point of view.

The supercharges $Q^1_-$, $\Qb_{2\dot-}$ and $Q^2_+$, $\Qb_{1\dot+}$
which comprise $Q$ and (part of) $\Qone$ form a subalgebra isomorphic
to the $\CN = (2,2)$ supersymmetry algebra on $\R^2$.  The latter may
be obtained by dimensional reduction from the $\CN = 1$ supersymmetry
algebra in four dimensions or the $\CN = 2$ supersymmetry algebra in
three dimensions.  It is generated by four supercharges $Q_\pm$,
$\Qb_\pm$, satisfying
\begin{equation}
  \begin{alignedat}{2}
  [Q_+, \Qb_+]
  &= P_w \,,
  &\qquad
  [Q_-, \Qb_-]
  &= -P_\wb \,.
  \\
  \label{eq:3dN=2SUSY-2}
  [Q_+, \Qb_-]
  &= Z \,,
  &
  [Q_-, \Qb_+]
  &= \Zb \,,
  \\
  [Q_+, Q_-]
  &=
  0
  \,,
  &
  [\Qb_+, \Qb_-]
  &=
  0 \,,
  \end{alignedat}
\end{equation}
where $Z$ is a complex central charge.  We have the identification
\begin{equation}
  Q^1_- = \Qb_- \,,
  \qquad
  \Qb_{2\dot-} = \Qb_+ \,,
  \qquad
  Q^2_+ = Q_+ \,,
  \qquad
  \Qb_{1\dot+} = -Q_- \,,
\end{equation}
and $Z$ coincides with the central charge in the four-dimensional
$\CN = 2$ supersymmetry algebra.  (In general, the $\CN = (2,2)$
supersymmetry algebra has an additional central charge, which is zero
in the present case.)

We may therefore think of any $\CN = 2$ supersymmetric field theory on
$\R^2 \times \C$ as an $\CN = (2,2)$ supersymmetric field theory on
$\R^2$.  From this two-dimensional point of view, a hypermultiplet
consists of a pair of $\CN = (2,2)$ chiral multiplets, which have
``continuous indices'' $(z,\zb)$, namely their coordinates on $\C$.

A chiral multiplet $\Phi$ of $\CN = (2,2)$ supersymmetry is the
dimensional reduction of a chiral multiplet of $\CN = 2$ supersymmetry
in three dimensions.  Under the action of
$\eps_- Q_+ - \eps_+ Q_- - \epsb_- \Qb_+ + \epsb_+ \Qb_-$, the fields
of $\Phi$ transform as
\begin{equation}
  \begin{aligned}
    \delta\phi
    &= \eps_-\psi_+ - \eps_+\psi_- \,,
    \\
    \delta\psi_+
    &= \iu\epsb_- \del_w\phi + \epsb_+ m\phi + \eps_+ F \,,
    \\    
    \delta\psi_-
    &= -\epsb_- \mb\phi + \iu\epsb_+ \del_\wb\phi + \eps_- F \,,
    \\    
    \delta F
    &=  -\iu\epsb_+(\del_\wb\psi_+ + \iu m\psi_-)
        - \iu\epsb_- (\iu \mb\psi_+ - \del_w\psi_-) \,,
  \end{aligned}
\end{equation}
with $m$ a complex mass parameter, the twisted mass of $\Phi$.  The
central charge is given by
\begin{equation}
  Z(\Phi) = m \,.
\end{equation}

Let us rename the fields of $\Phi$ and its conjugate $\Phib$ as
\begin{equation}
  \begin{alignedat}{2}
    \chi &= \iu\psi_+ \rmd w - \iu\psi_- \rmd\wb \,,
    &\qquad
    G &= F \, \rmd w \wedge \rmd\wb \,, \\
    \etab &= \psib_+ + \psib_- \,,
    &
    \mub &= \frac12(\psib_+ - \psib_-) \,.
  \end{alignedat}
\end{equation}
Then, the action of $Q = \Qb_- + \Qb_+$ on the fields can be written
as
\begin{equation}
  \begin{alignedat}{2}
    Q \cdot \phi
    &= 0 \,,
    &
    Q \cdot \phib
    &= \etab \,,
    \\
    Q \cdot \chi
    &= \rmd_m\phi \,,
    &
    Q \cdot \etab
    &= 0 \,,
    \\    
    Q \cdot G
    &= \rmd_m \chi \,,
    & \qquad
    Q \cdot \mub
    &= \Fb \,,
    \\    
    &&
    Q \cdot \Fb
    &= 0 \,,
  \end{alignedat}
\end{equation}
where
\begin{equation}
  \rmd_m = \rmd w (\del_w + \iu m) + \rmd\wb (\del_\wb - \iu\mb)
\end{equation}
is the exterior derivative twisted by the twisted mass $m$.  As in the
case of $\CN = 2$ supersymmetry in three dimensions, in terms of the
twisted-translated fields the twisted mass disappears from the formula.

We are interested in a theory consisting of multiple chiral multiplets
$\Phi^a = (\phi^a, \psi_\pm^a, F^a)$, coupled through a superpotential
$W$, which is a holomorphic function of $\phi^a$.  Integrating out the
auxiliary fields sets
\begin{equation}
  F^a = -g^{a\bb} \frac{\del\Wb}{\del\phib^\bb} \,,
  \qquad
  \Fb^\ab = -g^{\ab b} \frac{\del W}{\del\phi^b} \,,
\end{equation}
with $g^{a\bb} = g^{\ab b} = \delta_{ab}$.  For a generic
superpotential, the fields of the antichiral multiplets $\Phib^\ab$ do
not contribute to the $Q$-cohomology.  Then, the $Q$-cohomology is
spanned (at $w = 0$) by holomorphic functions of $\phi^a$, but there
is the relation
\begin{equation}
  \label{eq:dW=0}
  \rmd W = 0
\end{equation}
coming from $Q \cdot \mub^\ab$.  Modulo $\rmd\zb$, $\Qone$ acts on $\Phi^a$
as
\begin{equation}
  \Qone \cdot \phi^a = \chi^a \,,
  \qquad
  \frac12 \Qone \cdot \chi^a = G^a \,,
\end{equation}
from which it follows
\begin{equation}
  \begin{aligned}
    (\phi^a)^*
    =
    \phi^a + \chi^a + G^a \,.
  \end{aligned}
\end{equation}

A hypermultiplet on $\R^2 \times \C$ consists of a pair of chiral
multiplets with opposite masses $m$ and $-m$, whose scalar fields
$q^a(z,\zb)$, $a = 1$, $2$, have $(R,J') = (1,1/2)$ and are coupled by
the superpotential
\begin{equation}
  W \propto
  \int_C \rmd z \wedge \rmd\zb
  \,
  \eps_{ab} q^a \del_\zb q^b \,.
\end{equation}
The $Q$-cohomology of local operators is the algebra of holomorphic
functions of $q^a$, and by the relation \eqref{eq:dW=0} its classes
vary holomorphically on $\C$, as they should.

The $\lambda$-bracket has $(R,J') = (-2,-1)$, hence
$\lbracket{\lambda}{\ec{q^a}}{\ec{q^b}}$ is a $Q$-cohomology class of
$(R,J') = (0,0)$.  Furthermore, it should be invariant under the
$\SU(2)$ symmetry rotating $(q^1, q^2)$ as a doublet.  The only
possibility is a multiple of $\eps^{ab}$.  Let us show this
explicitly.

As explained near the end of section \ref{sec:PVA-4d}, we can compute
the $\lambda$-bracket by taking the integration cycle to be
$\R^2 \times S^1_{z_2}$:
\begin{equation}
  \lbracket{\lambda}{\ec{q^a}}{\ec{q^b}}(z_2)
  =
  \biggec{\biggl(\int_{\R^2 \times S^1_{z_2}}
  e^{\lambda (z_1 - z_2)} \rmd z_1 \wedge
  ({}^Zq^a)^*(x_1) \biggr) q^b(z_2)} \,.
\end{equation}
We have
\begin{equation}
  \biggl(\int_{\R^2} ({}^Zq^a)^*(x_1)\biggr) q^b(0)
  \propto
  \biggl(\int_{\R^2} e^{\iu m w_1 - \iu\mb\wb_1}
  g^{a\cb} \eps_{\cb\db} \del_z\qb^\db(x_1)
  \rmd w_1 \wedge \rmd\wb_1\biggr)
  q^b(x_2) \,,
\end{equation}
which, using the propagator, we can rewrite as
\begin{equation}
  \int_{\R^2}
  e^{\iu m w_1 - \iu\mb\wb_1}
  \eps^{ab}
  \del_{z_1} \biggl(\int_{\R^4} \rmd^4 p
  \frac{e^{-\iu p_\mu (x_1^\mu - x_2^\mu)}}{p^\mu p_\mu + |m|^2}\biggr)
  \rmd w_1 \wedge \rmd\wb_1 \,.
\end{equation}
Integrating over $w_1$, $\wb_1$ yields delta functions which set
$p_w = m$ and $p_\wb = -\mb$, leaving
\begin{equation}
  \eps^{ab}
  \del_{z_1}
  \biggl(
  \int_{\R^2} \rmd^2 p
  \frac{e^{-\iu p_z (z_1 - z_2) - \iu p_\zb(\zb_1 - \zb_2)}}{p_z p_\zb}
  \biggr)
  \propto
  \frac{\eps^{ab}}{z_1 - z_2} \,.
\end{equation}
The last proportionality can be seen from the fact that acted on with
$\del_{\zb_1}$, both sides become a delta function supported at
$z_1 = z_2$.  Performing the integral over $S^1_{z_2}$, we find
\begin{equation}
  \lbracket{\lambda}{\ec{q^a}}{\ec{q^b}}
  \propto
  \eps^{ab} \,,
\end{equation}
as expected.  The result is also independent of $m$ due to the
locality of $\CV$, as explained in section~\ref{sec:3d-free-chiral}.

For $m = 0$, the hypermultiplet is conformal.  The associated vertex
algebra is known to be the algebra of symplectic bosons
\cite{Beem:2013sza}, characterized by the OPE
\begin{equation}
  \ec{(q^a)^\hbar(z_1)} \ec{(q^b)^\hbar(z_2)}
  \sim
  \hbar \frac{\eps^{ab}}{z_1 - z_2} \,.
\end{equation}
In the classical limit $\hbar \to 0$, this vertex algebra indeed
reduces to the Poisson vertex algebra just found.

The Poisson algebra $\CV/(\del_z\CV) \CV$ is isomorphic to that of
holomorphic functions on $\C^2$ with respect to the holomorphic
symplectic form $\eps_{ab} \, \rmd q^a \wedge \rmd q^b$.  This is the
Poisson algebra associated with the Rozansky--Witten
twist~\cite{Rozansky:1996bq} of a three-dimensional $\CN = 4$
hypermultiplet~\cite{Beem:2018fng}.

\subsection{Gauge theories}

Finally, let us determine the Poisson vertex algebra for an $\CN = 2$
supersymmetric gauge theory, constructed from a vector multiplet for a
gauge group $G$ and a hypermultiplet in a representation $\rho$ of
$G$.

If $\rho$ is chosen appropriately and the hypermultiplet masses are
zero, the theory is conformal.  In this case the associated vertex
algebra can be defined, and it is known to be the algebra of gauged
symplectic bosons \cite{Beem:2013sza,Oh:2019bgz,Jeong:2019pzg}.
This algebra is constructed as follows.

Let $\gamma$ and $\beta$ be bosonic fields of conformal weight
$J' = 1/2$, valued in $\rho$ and its dual $\rho^\vee$, and $b$ and $c$
be fermionic fields with $J' = 1$ and $0$, valued in the adjoint
representation of $G$.  The ghost field $c$ is constrained to have no
zero mode.  The dynamics of these fields are governed by the action
\begin{equation}
  \frac{1}{\hbar} \int_C \rmd z \wedge \rmd\zb
  (b \del_\zb c + \beta \del_\zb \gamma) \,,
\end{equation}
which lead to the OPEs
\begin{equation}
  \gamma(z_1) \beta(z_2)
  \sim \frac{\hbar \id_\rho}{z_1 - z_2} \,,
  \qquad
  b(z_1) c(z_2)
  \sim \frac{\hbar C_2(\gf)}{z_1 - z_2} \,.
\end{equation}
Here $\id_\rho$ is the identity operator on the representation space
of $\rho$, and $C_2(\gf)$ is the quadratic Casimir element of the Lie
algebra $\gf$ of $G$.  Let $V_{\text{$bc$--$\beta\gamma$}}^\hbar$ be
the vertex algebra generated by $\beta$, $\gamma$, $b$, $c$.

The vertex algebra $V_{\text{$bc$--$\beta\gamma$}}^\hbar$ has a
fermionic symmetry, called the BRST symmetry, whose conserved current
is given by
\begin{equation}
  J_\BRST = \Tr(bcc) - \beta c \gamma \,.
\end{equation}
The corresponding charge $Q_\BRST$ acts on an element
$\CO \in V_{\text{$bc$--$\beta\gamma$}}^\hbar$ by
\begin{equation}
  \label{eq:QBRST-O}
  Q_\BRST \cdot \CO(z)
  =
  \frac{1}{2\pi\iu \hbar}
  \biggl(\int_{S^1_z} J_\BRST\biggr)
  \CO(z)
\end{equation}
and satisfies $Q_\BRST^2 = 0$.  The vertex algebra of the gauge theory
is the $Q_\BRST$-cohomology of $V_{\text{$bc$--$\beta\gamma$}}^\hbar$.
The vertex algebra for a free massless hypermultiplet corresponds to
the case when $G$ is trivial and
$(\ec{q_1}, \ec{q_2}) = (\gamma, \beta)$.

The Poisson vertex algebra associated with the gauge theory is the
classical limit of the BRST cohomology, and can be described as
follows.  The vertex algebra $V_{\text{$bc$--$\beta\gamma$}}^\hbar$
reduces to a Poisson vertex algebra $\CV_{\text{$bc$--$\beta\gamma$}}$
whose $\lambda$-bracket is given by
\begin{equation}
  \lbracket{\lambda}{\gamma}{\beta}
  \propto \id_\rho \,,
  \qquad
  \lbracket{\lambda}{b}{c}
  \propto C_2(\gf) \,.
\end{equation}
The action~\eqref{eq:QBRST-O} of $Q_\BRST$ on
$V_{\text{$bc$--$\beta\gamma$}}^\hbar$ is given by $1/\hbar$ times
$\lBracket{\lambda}{J_\BRST}{\ }|_{\lambda = 0}$.  In the limit
$\hbar \to 0$, this becomes the action of the zero mode
$\zmode J_\BRST$ on $\CV_{\text{$bc$--$\beta\gamma$}}$.  Therefore,
the Poisson vertex algebra for the $\CN = 2$ superconformal gauge
theory is the classical BRST cohomology of
$\CV_{\text{$bc$--$\beta\gamma$}}$, whose differential is given by
$\{\zmode J_\BRST, \ \}$.

The classical BRST cohomology of $\CV_{\text{$bc$--$\beta\gamma$}}$
actually makes sense for any choice of $\rho$, not necessarily one for
which the theory is conformal.  This is in contrast to its quantum
counterpart: $Q_\BRST$ squares to zero if and only if the one-loop
beta function vanishes.  The anomaly in the BRST symmetry arises from
double contractions in the OPE between two $J_\BRST$, which is of
order $\hbar^2$ and discarded in the classical limit.

Based on this observation, we propose that the Poisson vertex algebra
for an $\CN = 2$ supersymmetric gauge theory, whether conformal or
not, is given by the same classical BRST cohomology.

The character of this Poisson vertex algebra can be calculated by a
matrix integral formula.  In \cite{Cordova:2015nma}, this formula was
employed as a definition of the Schur index for nonconformal theories
with Lagrangian descriptions, and shown to coincide with a certain
wall-crossing invariant in a number of examples.

\section*{Acknowledgments}
  We would like to thank Kevin Costello for invaluable advice and
  illuminating discussions, and Dylan Butson, Tudor Dimofte and Davide
  Gaiotto for helpful comments.  The research of JO is supported in
  part by Kwanjeong Educational Foundation, by the Visiting Graduate
  Fellowship Program at the Perimeter Institute for Theoretical
  Physics, and by the Berkeley Center of Theoretical Physics.  The
  research of JY is supported by the Perimeter Institute for
  Theoretical Physics.  Research at Perimeter Institute is supported
  in part by the Government of Canada through the Department of
  Innovation, Science and Economic Development Canada and by the
  Province of Ontario through the Ministry of Colleges and
  Universities.

\providecommand{\href}[2]{#2}\begingroup\raggedright\endgroup

\end{document}